\documentclass[twocolumn,floatfix,preprintnumbers,superscriptaddress]{revtex4-1}

\usepackage[utf8]{inputenc}
\usepackage[colorlinks=true,citecolor=blue,linkcolor=blue]{hyperref}
\usepackage[normalem]{ulem}
\usepackage{url}
\usepackage{graphicx,wrapfig,float,slashed,cancel}
\usepackage{amsmath,amssymb,epsfig,graphicx,xcolor,stmaryrd}
\usepackage{bm}
\usepackage{enumitem}
\usepackage{subfigure}

\definecolor{darkblue}{RGB}{1, 90, 173}

\begin{document}   

\title{Properties of  spin-1/2 heavy baryons at nonzero temperature}
\date{\today}

\author{A. T{\"u}rkan}
\affiliation{\"Ozye\v{g}in University, Department~of Natural~and~Mathematical~Sciences, \c{C}ekmek\"{o}y, 34794 Istanbul, Turkey}
\author{G. Bozk{\i}r }
\affiliation{National Defense University, Army NCO Vocational HE School, Department of Basic Sciences, Alt{\i}eyl\"{u}l, 10185 Bal{\i}kesir ,Turkey}
\author{K.~Azizi}
\thanks{Corresponding Author}
\affiliation{University of Tehran, Department of Physics, North Karegar Avenue, Tehran,
14395-547 Iran}
\affiliation{Do\v{g}u\c{s} University, Department of Physics, Ac{\i}badem-Kad{\i}k\"{o}y, 34722
Istanbul, Turkey}

\begin{abstract}

The spectroscopic properties of single heavy spin-1/2 $\Lambda_{Q}$, $\Sigma_{Q}$, $\Xi^{(\prime)}_{Q}$  and $ \Omega_{Q}$ baryons  are investigated at finite temperature in the framework of the  thermal QCD sum rule.  We discuss the behavior of the mass and residue of these baryons with respect to temperature taking into account contributions of non-perturbative operators up to dimension eight. We  include  additional operators coming from the Wilson expansion due to breaking  the Lorentz invariance at nonzero temperature. The obtained results show that the mass of these baryons remain stable up to roughly  $T=108$ MeV while their residue is unchanged up to $T=93$ MeV. After these points, the mass and residue start to diminish by increasing the temperature. The shifts in the  mass and residue for both the bottom and charm channels are considerably large and we observe the melting of these baryons near to thepseudocritical temperature determined by recent lattice QCD calculations. We present our results for the mass of these baryons with both the positive and negative parity at the   $ T\rightarrow 0 $ limit, which are consistent with the existing theoretical predictions as well as experimental data.

\end{abstract}

\maketitle



%

\section{Introduction}  \label{sec:intro}

One of the most attractive subjects in particle physics is to investigate the spectroscopic properties of hadrons at finite temperatures. Such studies provide us with a better understanding of the perturbative and non-perturbative natures of QCD at hot mediums. They will  also help us in analyses of data provided by future heavy ion collision experiments aiming to investigate  the hadronic properties and possible phase transitions to quark gluon plasma (QGP) at finite temperatures and densities. In the last two decades, there have been significant experimental and theoretical studies on single heavy baryons in vacuum. Roughly, all single heavy baryons containing a heavy $ b $ or $ c $ quark  have been successfully observed \cite{PDG}.  The investigation of these baryons at a medium with nonzero temperature  is  a very  prominent research subject now and it will be in agenda of different experimental and theoretical studies.

Single heavy baryons $(Q q_{1} q_{2})$ are composite particles made of one heavy ($Q=b$ or  $c $) and two light quarks ($q_{1,2}=u, d$ or  $ s $). These particles belong to either antitriplet  of flavor antisymmetric state $  \overline{\textbf{3}} $ or sextet of flavor symmetric state $ \textbf{6}$. It is well known that total spin-parity of the ground state single heavy baryons in sextet state is either  $ J^{P}=\frac{1}{2}^{+} $ or $ J^{P}=\frac{3}{2}^{+} $  while spin-parity of the single heavy baryons in antitriplet state is only $ J^{P}=\frac{1}{2}^{+} $. In this paper, we study the spectroscopic parameters of the spin-1/2 heavy bottom/charmed baryons both in antitriplet and sextet representations, whose members are shown in Table \ref{tab:Baryons}.    

\begin{table}[ht!] 
	\centering
	\begin{tabular}{|c|c|c|c|}
\hline 
\,\, Baryon \,\,  & \,\, $q_{1}$ \,\, &  \,\, $q_{2}$ \,\,&  \,\, SU(3)  \,\,    \\[2mm]
\hline
 $\Lambda_{b(c)}^{0(+)}$ & $u$  & $d$   &
 $\bar 3$ \\
 $\Xi_{b(c)}^{0(+)}$ & $u$  & $s$   &
 $\bar 3$ \\
 $\Xi_{b(c)}^{-(0)}$ & $d$ & $s$   &
 $\bar 3$  \\
\hline
 $\Sigma_{b(c)}^{+(++)}$  & $u$  & $u$   &
$6$ \\
 $\Sigma_{b(c)}^{0(+)}$ & $u$  & $d$ &
$6$ \\
 $\Sigma_{b(c)}^{-(0)}$ & $d$   & $d$    &
$6$ \\
 $\Xi_{b(c)}^{0(+)'}$ & $u$  & $s$  &
$6$ \\
 $\Xi_{b(c)}^{-(0)'}$ & $d$ & $s$ &
$6$ \\
 $\Omega_{b(c)}^{-(0)}$ & $s$  & $s$ &
$6$ \\ 
\hline
\end{tabular}
	\caption{Quark content of the spin-1/2 heavy baryons with different charges.}
	\label{tab:Baryons}
\end{table}

In the literature, a lot of theoretical studies on the investigation of the spin-1/2 heavy baryons in vacuum have been performed using different phenomenological approaches such as 
the quark model \cite{Roberts,Karliner}, quark potential model \cite{Capstick}, 
heavy quark effective theory (HQET) \cite{Dai,Liu,Korner}, 
chiral perturbation theory \cite{Savage}, Feynman-Hellman theorem \cite{Roncaglia}, hypercentral approach \cite{Ghalenovi1,Ghalenovi2,Patel1,Patel2}, lattice QCD simulation \cite{Brown,Mathur,Lewis,Bahtiyar}, relativistic (constituent) quark model \cite{Ebert1,Ebert2,Ebert3,Migura,Gerasyuta1,Gerasyuta2,Gerasyuta3,Garcilazo}, chiral quark-soliton model \cite{Kim},  symmetry-preserving treatment of a vector$\times$vector contact interaction model \cite{Yin}, QCD sum rules \cite{Shuryak,Bagan,Azizi,Wang1,Wang2,Wang3,Zhang1,Zhang2,Agaev1,Agaev2,Agaev3,Agaev4},  etc. As we also previously mentioned, to better understand the hot and dense QCD matter created in relativistic heavy-ion collision experiments such as  Relativistic Heavy Ion Collider (RHIC) at Brookhaven National Laboratory’s (BNL) and the Large Hadron Collider (LHC) at the European Organization for Nuclear Research’s (CERN), the investigation of effects of temperature on spectroscopic properties of hadrons at nonzero temperature is needed. Such investigations  can also help us  to improve our understanding of phase transition, quark-gluon deconfinement and chiral symmetry restoration. By increasing in the  temperature, a transition or chiral crossover \cite{Aoki,MCheng} from the hadronic phase to QGP phase may  be occurred. Lattice QCD calculations show that thepseudocritical temperature is $ T_{pc}\approx155 MeV$ for chiral crossover \cite{Bhattacharya,Bazavov2} to QGP. 

One of the most applicable and powerful phenomenological tools that can be used to analyze the spectroscopic properties of hadrons at nonzero temperature is the thermal QCD sum rule method (TQCDSR). This method is the thermal version  \cite{Bochkarev} of the QCD sum rule, firstly introduced by Shifman, Vainshtein and Zakharov for mesons at zero temperature \cite{Shifman} and then applied to baryons in vacuum by Ioffe \cite{Ioffe}. In  thermal version, some additional operators appear in the operator product expansion (OPE/Wilson expansion) due to the breaking of the Lorentz invariance and vacuum expectation values are replaced by their thermal forms at finite temperature. The essential objective of this study is to extend our previous work on the thermal properties of the spin-3/2 heavy baryons at nonzero temperature \cite{AziziTurkan} and investigate  the shifts on the mass and residue of the spin-1/2 heavy $\Lambda_{Q}$, $\Xi_{Q}$, $\Sigma_{Q}$, $\Xi_{Q}^{'}$ and $ \Omega_{Q}$ baryons with respect to temperature using TQCDSR. In our calculations, we take account the extra operators arising from the OPE at nonzero temperature and use  the thermal quark, gluon and mixed condensates up to dimension eight as well as the temperature-dependent energy-momentum tensor.
      
This study is structured as follows: In Sec. II, we derive the TQCDSR for masses and residues of the spin-1/2 heavy baryons at nonzero temperature. In Sec. III, we present the numerical analysis of the obtained sum rules for the physical parameters and a comparison of our results at zero temperature with those existing in the literature. The last section is devoted to both the summary of  the results  and our concluding remarks.

\section{Thermal QCD Sum Rule Calculations }

\label{sec:DA} %

The aim of this section is to obtain TQCDSR for the masses and residues of the spin-1/2 heavy $\Lambda_{Q}$, $\Xi_{Q}$, $\Sigma_{Q}$, $\Xi_{Q}^{'}$ and $ \Omega_{Q}$ baryons at nonzero temperature. For this purpose, we start our calculations by considering the following two-point thermal correlation function:
\begin{equation}\label{CorrFunc}
\Pi(q,T)=i\int d^{4}x~e^{iq\cdot x}\langle
\Psi|\mathcal{T}\{J_{B_{Q}}(x) \bar{J}_{B_{Q}}(0)\}|\Psi\rangle,
\end{equation}
where  $q$ denotes the four-momentum of the considered spin-1/2 heavy baryon (${B_{Q}}$), $\Psi $ indicates the ground state of the hot medium and $\mathcal{T}$ is the time-ordering operator. $J_{B_{Q}}(x)$ is the interpolating current of  ${B_{Q}}$ baryon, which couples to  both the positive and negative parities.  It is represented by the following   expressions for anti-triplet ($\overline{\textbf{3}}$) and sextet ($\textbf{6}$) baryons \cite{Bagan}: 

\begin{eqnarray}
\label{currentantitriplet}
J_{\overline{\textbf{3}}}&=& \dfrac{1}{\sqrt{6}} \epsilon_{abc} \sum_{l=1}^2 \
\Big\{2\Big( q_1^{T,a}(x)C \Gamma_{1}^{l}q_2^{b}(x)\Big)\Gamma_{2}^{l} Q^c(x)\notag \\
&+&\Big( q_1^{T,a}(x)C \Gamma_{1}^{l}Q^b(x)\Big)\Gamma_{2}^{l}q_2^{c}(x)\notag \\
&+&\Big( Q^{T,a}(x)C \Gamma_{1}^{l}q_2^b(x)\Big)\Gamma_{2}^{l}q_1^{c}(x)\Big\},
\end{eqnarray}

\begin{eqnarray}
\label{currentsextet}
J_{\textbf{6}}&=&- \dfrac{1}{\sqrt{2}}\epsilon_{abc} \sum_{l=1}^2 \
\Big\{\Big( q_1^{T,a}(x)C \Gamma_{1}^{l}Q^{b}(x)\Big)\Gamma_{2}^{l} q_2^c(x)\notag \\
&-&\Big( Q^{T,a}(x)C \Gamma_{1}^{l}q_2^b(x)\Big)\Gamma_{2}^{l}q_1^{c}(x)\Big\},
\end{eqnarray}
where $ \epsilon_{abc} $ is the  Levi-Civita tensor with color indices $a, b, c$ , $C$ is the charge conjugation operator, $\Gamma_1^1 = 1$,~$\Gamma_1^2 = \Gamma_2^1 = \gamma^5$, and $\Gamma_2^2 =t$ in which $t$ denotes an arbitrary mixing parameter with  $t=-1$ corresponds to the famous Ioffe currents that we consider in the present study. As we previously noted, $q_{1}$ and $q_{2}$ stand for light quarks and  $Q$ for heavy quark field and they are given in Table  \ref{tab:Baryons} for all considered ${B_{Q}}$ baryons. Thus, considering Eq. (\ref{currentantitriplet}) and Eq. (\ref{currentsextet}), the interpolating currents for each state can be written as 
\begin{eqnarray}
\label{currents}
J_{\Xi_{Q}^{-(0)}}&=& -\dfrac{1}{\sqrt{6}} \epsilon_{abc} \sum_{l=1}^2 \
\Big\{2\Big( d^{T,a}(x)C \Gamma_{1}^{l}s^{b}(x)\Big)\Gamma_{2}^{l} Q^c(x)\notag \\
&+&\Big( d^{T,a}(x)C \Gamma_{1}^{l}Q^b(x)\Big)\Gamma_{2}^{l}s^{c}(x)\notag \\
&+&\Big( Q^{T,a}(x)C \Gamma_{1}^{l}s^b(x)\Big)\Gamma_{2}^{l}d^{c}(x)\Big\},\notag \\
J_{\Xi_{Q}^{0(+)}} &=& J_{\Xi_{Q}^{-(0)}}(d \rightarrow u),\notag \\
J_{\Lambda_{Q}^{0(+)}} &=& J_{\Xi_{Q}^{-(0)}}(d \rightarrow u, s \rightarrow d),\notag \\
J_{\Sigma_{Q}^{0(+)}}&=&- \dfrac{1}{\sqrt{2}}\epsilon_{abc} \sum_{l=1}^2 \
\Big\{\Big( u^{T,a}(x)C \Gamma_{1}^{l}Q^{b}(x)\Big)\Gamma_{2}^{l} d^c(x)\notag \\
&-&\Big( Q^{T,a}(x)C \Gamma_{1}^{l}d^b(x)\Big)\Gamma_{2}^{l}u^{c}(x)\Big\},\notag \\
J_{\Xi_{Q}^{-(0)'}} &=& J_{\Sigma_{Q}^{0(+)}}(u \rightarrow d, d \rightarrow s),\notag \\
J_{\Xi_{Q}^{0(+)'}} &=& J_{\Xi_{Q}^{-(0)'}}(d \rightarrow u),\notag \\
J_{\Sigma_{Q}^{-(0)}} &=& J_{\Xi_{Q}^{-(0)'}}(s \rightarrow d),\notag \\
J_{\Sigma_{Q}^{+(++)}} &=& J_{\Sigma_{Q}^{-(0)}}(d \rightarrow u),\notag \\
J_{\Omega_{Q}^{-(0)}} &=& J_{\Sigma_{Q}^{+(++)}}(u \rightarrow s).
\end{eqnarray}
According to the standard philosophy of the QCD sum rule method, the aforementioned thermal correlation function is evaluated in two basic ways: i) On the hadronic side, it is calculated in terms of hadronic parameters such as the temperature-dependent mass and residue of hadron. ii) On the QCD side, it is calculated  in terms  of temperature-dependent QCD degrees of freedom in the deep Euclidean region with the help of OPE. Then,  matching the coefficients of the selected structures  from both sides in momentum space, via the dispersion relation, the QCD sum rules for the spectroscopic parameters of the ${B_{Q}}$ at nonzero temperature are obtained. In the final step, to suppress the contributions of the higher states and continuum in obtained sum rules, Borel transformation and continuum subtraction are  applied to both sides of these equalities. One may first apply the Borel transformation and continuum subtraction, then match the coefficients of the selected structures from both sides.

The thermal correlation function in the hadronic side is obtained by inserting the full set of hadronic states having the same quantum numbers as the related interpolating current into Eq. (\ref{CorrFunc}). After performing the  integration over four-$x$, the thermal correlation function for the hadronic side can be written as 

\begin{eqnarray}\label{piphys}
&&\Pi^{Had.}(q,T)\notag \\&=&-\frac{{\langle}\Psi|J_{B_{Q}}(0)|B_{Q}^{+}(q,s){\rangle}
{\langle}B_{Q}^{+}(q,s)|J^{\dag}_{B_{Q}}(0)|\Psi{\rangle}}{q^{2}-m_{B_{Q}^{+}}^{2}(T)} \notag \\
&-&\frac{{\langle}\Psi|J_{B_{Q}}(0)|B_{Q}^{-}(q,s){\rangle}
{\langle}B_{Q}^{-}(q,s)|J^{\dag}_{B_{Q}}(0)|\Psi{\rangle}}{q^{2}-m_{B_{Q}^{-}}^{2}(T)} \notag \\
&+&\mbox{...},\notag \\
\end{eqnarray}
where  $|B_{Q}^{+}(q,s)\rangle$ and $|B_{Q}^{-}(q,s)\rangle$ are spin-1/2 single heavy baryon states with the positive and negative parity, respectively,  dots stand for the contributions of the higher states and continuum and $m_{B_{Q}^{\pm}}(T)$ is the temperature-dependent mass of $ B_{Q}^{\pm} $. The matrix elements  ${\langle}\Psi|J_{B_{Q}}(0)|B_{Q}^{\pm}(q,s){\rangle}$ for $B_{Q}^{\pm}$ are defined as

\begin{eqnarray}\label{matrixelement}
{\langle}\Psi|J_{B_{Q}}(0)|B_{Q}^{+}(q,s){\rangle}&=&\lambda_{B_{Q}^{+}}(T)u_{B_{Q}^{+}}(q,s),\nonumber \\
{\langle}\Psi|J_{B_{Q}}(0)|B_{Q}^{-}(q,s){\rangle}&=&\lambda_{B_{Q}^{-}}(T)\gamma_5 u_{B_{Q}^{-}}(q,s),
\end{eqnarray}
where $\lambda_{B_{Q}^{\pm}}(T)$ is the temperature-dependent residue of $B_{Q}^{\pm}$ and $u_{B_{Q}^{\pm}}(q,s)$ is Dirac spinor of spin $s$ and the four-momentum $q$. To proceed, we insert Eq. (\ref{matrixelement}) into Eq. (\ref{piphys}) and perform summation over spins of $B_{Q}^{\pm}$. The hadronic side of thermal correlation function in its final form  is decomposed  in terms of different structures  as
\begin{eqnarray}
\label{}
\Pi^{Had.}(q,T)&=&-\frac{\lambda^{2}_{B_{Q}^{+}}(T)(\!\not\!{q}+m_{B_{Q}^{+}}(T))}{q^{2}-m^{2}_{B_{Q}^{+}}(T)}\nonumber \\
&-&\frac{\lambda^{2}_{B_{Q}^{-}}(T)(-\!\not\!{q}+m_{B_{Q}^{-}}(T))}{q^{2}-m^{2}_{B_{Q}^{-}}(T)}+\cdots.
\end{eqnarray}
This correlation function can be written in terms of the structures $\!\not\!{q}$ and $I$ as 
\begin{eqnarray}
\Pi^{Had.}(q,T)&=&\Pi_{1}^{Had.}(T)\!\not\!{q}+\Pi_{2}^{Had.}(T)I+\cdots, \notag \\
\end{eqnarray}
where $\Pi^{Had.}_{1(2)}(T)$,  as the coefficients of the selected Lorentz structures,   in Borel scheme are  obtained as 
\begin{eqnarray}\label{Physical1}
\hat{B}\Pi_{1}^{Had.}(T)&=&
 \lambda^{2}_{B_{Q}^{+}}(T)e^{-m_{B_{Q}^{+}}^{2}(T)/M^2}\nonumber \\
&-& \lambda^{2}_{B_{Q}^{-}}(T)e^{-m_{B_{Q}^{-}}^{2}(T)/M^2},
\end{eqnarray}
and 
\begin{eqnarray}\label{Physical2}
\hat{B}\Pi_{2}^{Had.}(T)&=&
 \lambda^{2}_{B_{Q}^{+}}(T)m_{B_{Q}^{+}}(T)e^{-m_{B_{Q}^{+}}^{2}(T)/M^2}\nonumber \\
&+&\lambda^{2}_{B_{Q}^{-}}(T)m_{B_{Q}^{-}}(T)e^{-m_{B_{Q}^{-}}^{2}(T)/M^2},
\end{eqnarray}
where $M^2$ is the Borel parameter to be determined  in next  section.

Now, we have to evaluate the QCD side of the thermal correlation function in terms of the quark-gluon degrees of freedom in the deep Euclidean region. For this aim, we insert the related interpolating current of ${B_{Q}}$ given in Eq. (\ref{currents}) into Eq. (\ref{CorrFunc}) and contract all light and heavy quark fields using the Wick theorem. Thus, the most general form of the thermal correlation function on the QCD side in terms of thermal light (heavy) quark propagators $S_{q(Q)}^{ij}(x)$ for $\overline{\textbf{3}}$  and $\textbf{6}$ baryons are  obtained as 

\begin{eqnarray}\label{antisymmetricq1difq2}
\Pi_{\overline{\textbf{3}}}^{QCD}(q,T)&=& \frac{i}{6}\epsilon_{abc}\epsilon_{a'b'c'} \int d^4 x e^{iq\cdot x}\notag \\&\times &\sum_{l=1}^2 \sum_{k=1}^2\
\left\lbrace \Gamma^{l}_{2} \Big(2 S^{ca'}_{Q}(x) A^{k}_{1}\widetilde{S}^{ab'}_{q_{1}}(x) \Gamma^{l}_{1} S^{bc'}_{q_{2}}(x) \Gamma^{k}_{2}\right. \notag \\
&+&2 S^{cb'}_{Q}(x) \Gamma^{k}_{1}\widetilde{S}^{ba'}_{q_{2}}(x) \Gamma^{l}_{1} S^{ac'}_{q_{1}}(x) \Gamma^{k}_{2}\notag \\&-&S^{ca'}_{q_{2}}(x) \Gamma^{k}_{2}\widetilde{S}^{bb'}_{Q}(x) \Gamma^{l}_{1} S^{ac'}_{q_{1}}(x) \Gamma^{k}_{1}\notag \\&-&2S^{ca'}_{q_{2}}(x) \Gamma^{k}_{2}\widetilde{S}^{ab'}_{q_{1}}(x) \Gamma^{l}_{1} S^{bc'}_{Q}(x) \Gamma^{k}_{1}\notag \\
&-&S^{cb'}_{q_{1}}(x) \Gamma^{k}_{2}\widetilde{S}^{aa'}_{Q}(x) \Gamma^{l}_{1} S^{bc'}_{q_{2}}(x) \Gamma^{k}_{1}\notag \\
&-&2S^{cb'}_{q_{1}}(x) \Gamma^{k}_{2}\widetilde{S}^{ba'}_{q_{2}}(x) \Gamma^{l}_{1} S^{ac'}_{Q}(x) \Gamma^{k}_{1}\Big)\notag \\&-&\Gamma^{k}_{2} \Big(S^{cc'}_{q_{1}}(x) \Gamma^{l}_{2}Tr [\Gamma^{k}_{1} S^{ab'}_{Q}(x) \Gamma^{l}_{1}\widetilde{S}^{ba'}_{q_{2}}(x)]\notag \\
&+&S^{cc'}_{q_{2}}(x)\Gamma^{l}_{2}
Tr [\Gamma^{k}_{1}S^{ab'}_{q_{1}}(x)\Gamma^{l}_{1} \widetilde{S}^{ba'}_{Q}(x)]\notag \\&+&4S^{cc'}_{Q}(x)\Gamma^{l}_{2}
Tr [\Gamma^{k}_{1}S^{ab'}_{q_{1}}(x)\Gamma^{l}_{1} \widetilde{S}^{ba'}_{q_{2}}(x)]\Big)\left. \right\rbrace ,
\end{eqnarray}

\begin{eqnarray}\label{symmetricq1difq2}
\Pi_{\textbf{6}}^{QCD}(q,T) &=& -\frac{i}{2}\epsilon_{abc}\epsilon_{a'b'c'} \int d^4 x e^{iq\cdot x}\notag \\&\times &\sum_{l=1}^2 \sum_{k=1}^2\
\left\lbrace \Gamma^{l}_{2} \Big(S^{ca'}_{q_{1}}(x) A^{k}_{1}\widetilde{S}^{ab'}_{Q}(x) \Gamma^{l}_{1} S^{bc'}_{q_{2}}(x) \right. \notag \\
&+&S^{cb'}_{q_{2}}(x) \Gamma^{k}_{1}\widetilde{S}^{ba'}_{Q}(x) \Gamma^{l}_{1} S^{ac'}_{q_{1}}(x)\Big)\Gamma^{k}_{2}\notag \\&+&\Gamma^{k}_{2}\Big(S^{cc'}_{q_{1}}(x) \Gamma^{l}_{2}Tr [\Gamma^{l}_{1} \widetilde{S}^{aa'}_{Q}(x)\Gamma^{k}_{1}S^{bb'}_{q_{2}}(x)]\notag \\
&+& S^{cc'}_{q_{2}}(x) \Gamma^{l}_{2}Tr [\Gamma^{l}_{1} \widetilde{S}^{aa'}_{q_{1}}(x)\Gamma^{k}_{1}S^{bb'}_{Q}(x)]\Big)\left.\right\rbrace,
\end{eqnarray}
where $\widetilde{S}^{ij}_{q(Q)}=CS^{ijT}_{q(Q)}C$. To proceed, thermal light (heavy) quark propagators $S_{q(Q)}^{ij}(x)$ in coordinate space are needed, which are used  as (see also  \cite{Azizi1, Azizi2, Prop_C})
\begin{eqnarray}\label{lightquarkpropagator}
S_{q}^{ij}(x)&=& i\frac{\!\not\!{x}}{ 2\pi^2 x^4}\delta_{ij}-\frac{m_q}{4\pi^2 x^2}\delta_{ij}-\frac{\langle\bar{q}q\rangle_{T}}{12}\delta_{ij} \notag \\
&-&\frac{ x^{2}}{192} m_{0}^{2}\langle
\bar{q}q\rangle_{T}\Big[1-i\frac{m_q}{6}\!\not\!{x}\Big]\delta_{ij}\nonumber\\
&+&\frac{i}{3}\Big[\!\not\!{x}\Big(\frac{m_q}{16}\langle
\bar{q}q\rangle_{T}-\frac{1}{12}\langle u^{\mu}\Theta_{\mu\nu}^{f}u^{\nu}\rangle\Big)\nonumber\\
&+&\frac{1}{3}\Big(u\cdot x\!\not\!{u}\langle
u^{\mu}\Theta_{\mu\nu}^{f}u^{\nu}\rangle\Big)\Big]\delta_{ij}
\nonumber\\
&-&\frac{ig_s \lambda_{A}^{ij}}{32\pi^{2} x^{2}}
G_{\mu\nu}^{A}\Big(\!\not\!{x}\sigma^{\mu\nu}+\sigma^{\mu\nu}
\!\not\!{x}\Big)\nonumber\\
&-&i\frac{x^2 \!\not\!{x} g_{s}^{2} \langle
	\bar{q}q\rangle_{T}^{2}}{7776}\delta_{ij}-\frac{x^4 \langle
	\bar{q}q\rangle_{T} \langle
	g_{s}^{2} G^2\rangle_{T}}{27648}+...~,\nonumber\\
\end{eqnarray}

\begin{eqnarray}\label{heavypropagator}
S_{Q}^{ij}(x)&=&i \int\frac{d^4k e^{-ik \cdot x}}{(2\pi)^4} 
\left( \frac{\!\not\!{k}+m_Q}{k^2-m_Q^2}\delta_{ij}\right.\nonumber\\
&-&\frac{g_{s}G^{\alpha\beta}_{i j}}{4} \frac{\sigma^{\alpha\beta}(\!\not\!{k}+m_Q)+(\!\not\!{k}+m_Q)\sigma^{\alpha\beta} }{(k^2-m_Q^2)^{2}} 
\nonumber\\
&+&\frac{ m_{Q}}{12}\frac{k^{2}+m_{Q}\!\not\!{k}}{(k^{2}-m_{Q}^{2})^{4}} \langle g_{s}^{2} G^2 \rangle_{T}\delta_{ij}+\cdots  \Bigg) . 
\end{eqnarray}  
Here, $m_{q(Q)}$  is the light (heavy) quark mass, $\langle\bar{q}q\rangle_{T}$ is the thermal quark condensate, $\langle g_{s}^{2} G^2\rangle_{T}$ is thermal gluon condensate,  $ m_{0}^{2}\langle\bar{q}q\rangle_{T}=\langle \bar{q}g_{s}\sigma Gq\rangle_{T}$ is the thermal mixed condensate. The new operators, emerging in OPE,  appear to restore the Lorentz invariance broken out by the choice of the thermal rest frame at nonzero temperature. They are expressed in terms of fermionic and gluonic parts of the energy momentum tensor $\Theta^{f,g}_{\mu\nu}$ $(  \langle u^\mu \Theta^{f,g}_{\mu\nu} u^\nu \rangle = \langle u \Theta^{f,g} u \rangle = \langle    \Theta^{f,g} _{00} \rangle = \langle  \Theta^{f,g}  \rangle )$ and the four-vector velocity of the medium, $u^{\mu}$. To this end,  $u^{\mu}=(1,0,0,0)$ is chosen which leads to $u^2=1$. In the rest frame of the heat bath,  $ q\cdot u=q_{0} $, with $q_{0} $ being the energy of quasi particle,  $q_{0}=E_{\vec{q}}=(\vert\vec{q}\vert^{2}+m^{2})^{1/2}$, in the mass-shell condition. At  $\vec{q}=0$  limit it is the mass of the particle.  In the light quark propagator, the fermionic part of the energy momentum tensor,  $\langle
u^{\mu}\Theta_{\mu\nu}^{f}u^{\nu}\rangle$,  is seen explicitly whereas the gluonic part, $\langle u^{\lambda} {\Theta}^g _{\lambda \sigma} u^{\sigma}\rangle$, appears in the trace of two-gluon field strength tensor in the heat bath \cite{Mallik}, i.e.

\begin{eqnarray}\label{TrGG} 
\langle Tr^c G_{\alpha \beta} G_{\mu \nu}\rangle &=& \frac{1}{24} (g_{\alpha \mu} g_{\beta \nu} -g_{\alpha
	\nu} g_{\beta \mu})\langle G^2\rangle_{T} \nonumber \\
&+&\frac{1}{6}\Big[g_{\alpha \mu}g_{\beta \nu} -g_{\alpha \nu} g_{\beta \mu} -2(u_{\alpha} u_{\mu}g_{\beta \nu} \nonumber \\
&-&u_{\alpha} u_{\nu} g_{\beta \mu} -u_{\beta} u_{\mu}
g_{\alpha \nu} +u_{\beta} u_{\nu} g_{\alpha \mu})\Big]\nonumber \\
&\times&\langle u^{\lambda} {\Theta}^g _{\lambda \sigma} u^{\sigma}\rangle.
\end{eqnarray}

The QCD side of the correlation function can be written similar to hadronic side in terms of different Lorentz structures as
\begin{eqnarray}
\Pi^{QCD}(q,T)&=&\Pi_{1}^{QCD}(T)\!\not\!{q}+\Pi_{2}^{QCD}(T) I,\notag \\
\end{eqnarray}
where $\Pi^{QCD}_{1(2)}(T)$ are the coefficients of the selected Lorentz structures and they  contain both the perturbative and non-perturbative contributions.  The perturbative and some non-perturbative  parts are written  in terms of  the dispersion integrals in the present study.  Thus,
\begin{equation}
\label{disp_0}
\Pi^{QCD}_{1(2)}(T)=\int_{s_{min}}^\infty ds \dfrac{\rho^{{QCD}}_{1(2)} (s,T)}{s-q^2}+\Gamma_{1(2)}^{QCD}(T),
\end{equation}
where $ s_{min}=(m_{q_{1}}+m_{q_{2}}+m_{Q})^{2} $,  $\rho^{ QCD}_{1(2)}(s,T)$ are the spectral densities and $\Gamma_{1(2)}^{QCD}(T)$ stand for the remaining non-perturbative contributions that are calculated directly by applying the Borel transformation. The related spectral densities are defined  as
\begin{equation}
\rho^{QCD}_{1(2)}(s,T)=\frac{1}{\pi}\mathrm{Im}[\Pi^{QCD}_{1(2)}(T)].
\end{equation}
After Borel transformation and continuum subtraction we get 
\begin{eqnarray}\label{Gammafunc}
\hat{B}\Pi_{1(2)}^{QCD} (T)=\int_{s_{min}}^{s_{0}(T)} ds \rho_{1(2)}^{{ QCD}} (s,T)e^{-s/M^2}+\hat{B}\Gamma_{1(2)}^{QCD}(T),\nonumber\\
\end{eqnarray}
where $ s_{0}(T) $ is the temperature-dependent continuum threshold.  The main task in this part is to find the expressions for  $\rho^{ QCD}_{1(2)}(s,T)$ and  $\hat{B}\Gamma_{1(2)}^{QCD}(T)$.
They are obtained after inserting the explicit forms of the heavy and light quark propagators into the QCD side of the thermal correlation function given in Eqs. (\ref{antisymmetricq1difq2})       
and (\ref{symmetricq1difq2}) for $\overline{\textbf{3}}$  and $\textbf{6}$ baryons with spin-1/2, performing  the Fourier integral to go to the momentum space and applying the steps above to get the perturbative and non-perturbative parts.  As a result, the explicit forms of the functions  $\rho^{ QCD}_{1(2)}(s,T)$ and  $\hat{B}\Gamma_{1(2)}^{QCD}(T)$ are obtained.  Because of obtained expressions are quite lengthy, we present the expressions of $\rho^{QCD}_{1}(s,T)$ and $\hat{B}\Gamma_{1}^{QCD}(T)$ for only $\Sigma_{b}^{0}$ baryon as an example in the Appendix.

Finally,  we match the coefficients of the selected structures from the hadronic and QCD sides of the correlation function and find the  sum rules:

 \begin{eqnarray}\label{residuesumrule}
 \hat{B}\Pi_{1}^{QCD}(T)&=&\lambda^{2}_{B_{Q}^{+}}(T)e^{-m_{B_{Q}^{+}}^{2}(T)/M^2}\nonumber \\
&-&\lambda^{2}_{B_{Q}^{-}}(T)e^{-m_{B_{Q}^{-}}^{2}(T)/M^2},
 \end{eqnarray}
 and
 \begin{eqnarray}\label{residuesumrule2}
 \hat{B}\Pi_{2}^{QCD} (T)&=& \lambda^{2}_{B_{Q}^{+}}(T)m_{B_{Q}^{+}}(T)e^{-m_{B_{Q}^{+}}^{2}(T)/M^2}\nonumber \\
&+&\lambda^{2}_{B_{Q}^{-}}(T)m_{B_{Q}^{-}}(T)e^{-m_{B_{Q}^{-}}^{2}(T)/M^2}.
 \end{eqnarray}
In order to obtain the four unknowns, $m_{B_{Q}^{+}}$, $m_{B_{Q}^{-}}$, $\lambda_{B_{Q}^{+}}$ and $\lambda_{B_{Q}^{-}}$, two more equations which can be achieved by applying derivatives with respect to  $ \frac{d}{d(-\frac{1}{M^2})} $ to both sides of Eqs. (\ref{residuesumrule}) and (\ref{residuesumrule2}) are needed. Therefore, we get 

\begin{eqnarray}\label{derivativeresiduesumrule}
\frac{d \Pi_{1}^{QCD}(T)}{d(-1/M^2)} &=&\lambda^{2}_{B_{Q}^{+}}(T)m_{B_{Q}^{+}}^{2}(T)e^{-m_{B_{Q}^{+}}^{2}(T)/M^2}\nonumber \\
&-&\lambda^{2}_{B_{Q}^{-}}(T)m_{B_{Q}^{-}}^{2}(T)e^{-m_{B_{Q}^{-}}^{2}(T)/M^2},
 \end{eqnarray}
\begin{eqnarray}\label{derivativeresiduesumrule2}
\frac{d \Pi_{2}^{QCD}(T)}{d(-1/M^2)} &=&\lambda^{2}_{B_{Q}^{+}}(T)m_{B_{Q}^{+}}^{3}(T)e^{-m_{B_{Q}^{+}}^{2}(T)/M^2}\nonumber \\
&+&\lambda^{2}_{B_{Q}^{-}}(T)m_{B_{Q}^{-}}^{3}(T)e^{-m_{B_{Q}^{-}}^{2}(T)/M^2},
 \end{eqnarray}
By simultaneously solving the  above four equations, the temperature-dependent masses and residues for the positive and negative parity spin-1/2 heavy baryons are obtained  in terms of $s_0(T)$, $M^2$, QCD degrees of freedom and other inputs. For the   masses,  as examples, we get

\begin{widetext}

\begin{eqnarray}\label{masspositiveparity}
m_{B_{Q}^{+}}=\dfrac{(\alpha_{4}\alpha_{1}-\alpha_{3}\alpha_{2}+\sqrt{4\alpha_{3}^{3}\alpha_{1}+\alpha_{4}^{2}\alpha_{1}^{2}-6\alpha_{3}\alpha_{4}\alpha_{1}\alpha_{2}-3\alpha_{3}^{2}\alpha_{2}^{2}+4\alpha_{4}\alpha_{2}^{3}})}{2\alpha_{3}\alpha_{1}-2\alpha_{2}^{2}},
 \end{eqnarray}

\begin{eqnarray}\label{massnegativeparity}
m_{B_{Q}^{-}}=\dfrac{(-\alpha_{4}\alpha_{1}+\alpha_{3}\alpha_{2}+\sqrt{4\alpha_{3}^{3}\alpha_{1}+\alpha_{4}^{2}\alpha_{1}^{2}-6\alpha_{3}\alpha_{4}\alpha_{1}\alpha_{2}-3\alpha_{3}^{2}\alpha_{2}^{2}+4\alpha_{4}\alpha_{2}^{3}})}{2\alpha_{3}\alpha_{1}-2\alpha_{2}^{2}},
 \end{eqnarray}
where

\begin{eqnarray}\label{constants}
\alpha_{1}= \hat{B}\Pi_{1}^{QCD}(T),   \alpha_{2}= \hat{B}\Pi_{2}^{QCD}(T),    \alpha_{3}=\frac{d \Pi_{1}^{QCD}(T)}{d(-1/M^2)},    \alpha_{4}=\frac{d \Pi_{2}^{QCD}(T)}{d(-1/M^2)}.   
 \end{eqnarray}

\end{widetext}

Similar results are obtained for the  temperature-dependent  residues.

\section{Numerical results}

In this section, we perform the numerical analyses of the sum rules for the masses and residues of the spin-1/2 heavy $\Lambda_{Q}$, $\Xi_{Q}$, $\Sigma_{Q}$, $\Xi_{Q}^{'}$ and $ \Omega_{Q}$ baryons at nonzero temperature. For this aim, firstly we use  the numerical values of some input parameters collected in Table \ref{tab:Param} in our calculations. 

\begin{table}[ht!] 
	\centering
	\begin{tabular}{ |c|c|}
		\hline \hline
		Parameters  &  Values   \\ \hline
		$q_0^{\Lambda_{b}} $; $ q_0^{\Lambda_{c}} $   &  $ (5619.6\pm0.17) $; $ (2286.46\pm0.14)  $ $MeV$    \\    
		$ q_0^{\Xi_{b}} $; $q_0^{\Xi_{c}} $   &  $ (5791.9\pm0.5) $; $ (2467.71\pm0.23)  $ $MeV$      \\  
		$ q_0^{\Sigma_{b}} $; $ q_0^{\Sigma_{c}}$   &  $(5810.56 \pm0.25) $;  $ (2452.9 \pm0.4)$  $MeV$  \\ 
		$ q_0^{\Xi_{b}^{'}} $; $ q_0^{\Xi_{c}^{'}}$   &  $( 5935.02 \pm0.02 \pm0.05) $;  $ ( 2578.7\pm0.5) $  $MeV$  \\ 
		$ q_0^{\Omega_{b}} $; $ q_0^{\Omega_{c}} $   &  $( 6046.1\pm1.7) $;  $ (2695.2\pm1.7)  $  $MeV$  \\ 
		$ m_{u}   $ ;  $ m_{d}   $       &  $(2.3_{-0.5}^{+0.7})$ $MeV$; $(4.8_{-0.3}^{+0.7})$ $MeV$   \\
		$  m_{s}   $          &  $(93_{-5}^{+11} )$ $MeV$      \\
		$ m_{b}   $ ; 	$ m_{c}   $         &   $(4.18_{-0.03}^{+0.04})$ $GeV$; $(1.275_{-0.035}^{+0.025})$ $GeV$   \\
		$m_{0}^{2};   $          &  $(0.8\pm0.2)$ $GeV^2$     \\
		$ \langle0|\overline{q}q|0\rangle  (q=u, d)$          &  $-(0.24\pm0.01)^3$ $GeV^3$        \\
		$ \langle0|\overline{s}s|0\rangle $          &  $-0.8(0.24\pm0.01)^3$ $GeV^3$         \\
		$ {\langle}0\mid \frac{1}{\pi}\alpha_s G^2 \mid 0{\rangle}$          &  $ 0.012(3)~GeV^4$  \\
		\hline \hline
	\end{tabular}
	\caption{Numerical values for  the energy of quasi particles in medium,  quark masses  and vacuum condensates \cite{Belyaev,Dosch,Ioffe1,PDG,Gubler}. In the rest frames of the heat bath and the particle, we set the energy of the quasi particle to its ground state  positive parity mass value at each channel. }
	\label{tab:Param}
\end{table}

To go further in the analyses, we also need to know the  thermal quark condensates $\langle\bar{q}q\rangle_{T}$ (for $ q=u, d$ and $s $),  parametrized in terms of vacuum condensates and temperature. For this purpose, we use the following parametrization, which are   based on lattice QCD results  presented in Ref. \cite{Gubler}  
\begin{eqnarray}\label{qbarq}
\langle\bar{q}q\rangle_{T}&=&(A e^{\frac{T}{B[GeV]}}+C)\langle0|\bar{q}q|0\rangle,
\end{eqnarray}
where the coefficients A, B and C for the corresponding $q=u,d$ and $s$ are given in Table \ref{tab:coefficients1}.  Note, that the lattice results in Ref. \cite{Gubler}   are given in a wide range of the temperature, however, we find their fit functions up to the critical temperature under consideration in the present study (see also \cite{AziziTurkan}). The above fit together with the parameters in the Table \ref{tab:coefficients1} exactly reproduce the graphics for the temperature-dependent quark condensates in   Ref. \cite{Gubler} .
\begin{table}[ht!] 
	\centering
	\begin{tabular}{ |c|c|c|c|}
		\hline \hline 
		    &  A   &    B [GeV]   &   C    \\ \hline
		for $ q=u, d $   &  $-6.534\times10^{-4}$  &    0.025    &    1.015   \\   
		for $ q=s $   &  $-2.169\times10^{-5}$  &    0.019    &    1.002   \\
		\hline \hline 
	\end{tabular}
	\caption{The coefficients A, B and C in the thermal quark condensates $\langle\bar{q}q\rangle_{T}$.}
	\label{tab:coefficients1}
\end{table}

The temperature-dependent gluon condensate $ \langle G^2\rangle_{T} $ is given as \cite{Gubler}: 
\begin{eqnarray}\label{G2TLattice}
\delta \langle \frac{\alpha_{s}}{\pi}G^{2}\rangle_{T}&=&-\frac{8}{9}[ \delta T^{\mu}_{\mu}(T)-m_{u} \delta \langle\bar{u}u\rangle_{T}\nonumber \\ &-&m_{d} \delta \langle\bar{d}d\rangle_{T}-m_{s} \delta \langle\bar{s}s\rangle_{T}],
\end{eqnarray}
where
\begin{eqnarray}\label{deltafT}
\delta f(T)\equiv f(T)-f(0),
\end{eqnarray}
and $ \delta T^{\mu}_{\mu}(T) $ is defined as
\begin{eqnarray}\label{deltaTmunu}
\delta T^{\mu}_{\mu}(T)=\varepsilon(T)-3p(T),
\end{eqnarray}
with $\varepsilon(T)$ being the energy density and $ p(T) $ is the pressure. Using the recent Lattice calculations given in \cite{Bazavov1,Borsanyi},  we obtain the following fit function for $ \delta T^{\mu}_{\mu}(T) $ (see also  \cite{AziziTurkan}):
\begin{eqnarray}\label{epsmines3p}
\delta T^{\mu}_{\mu}(T)&=&(0.020 e^{\frac{T}{0.034[GeV]}}+0.115)T^{4}.
\end{eqnarray}
Note that this function, obtained in the present study, exactly reproduces the lattice QCD graphics for $\delta T^{\mu}_{\mu}(T) $ with respect to temperature presented in Refs. \cite{Bazavov1,Borsanyi}. 
The  temperature-dependent strong coupling is also given as \cite{Kaczmarek,Morita} 
\begin{eqnarray}\label{geks2T}
g_s^{-2}(T)=\frac{11}{8\pi^2}\ln\Big(\frac{2\pi
	T}{\Lambda_{\overline{MS}}}\Big)+\frac{51}{88\pi^2}\ln\Big[2\ln\Big(\frac{2\pi
	T}{\Lambda_{\overline{MS}}}\Big)\Big],\nonumber \\
\end{eqnarray}
where, $\Lambda_{\overline{MS}}\simeq T_{pc}/1.14$.

We use results on the thermal behavior of the energy-momentum tensor given by lattice QCD in Ref. \cite{Bazavov1} and parametrize the gluonic and fermionic parts of the energy density up to thepseudocritical temperature. Hence, we use: 
\begin{eqnarray}\label{tetaf}
\langle\Theta^{f(g)}\rangle&=& (D e^{\frac{T}{E[GeV]}}+F)T^{4},
\end{eqnarray}
with the related coefficients defined in Table \ref{tab:coefficients2}. This function, obtained in the present study, reproduces the temperature-dependent energy densities presented in Ref. \cite{Bazavov1} by graphics.
\begin{table}[ht!] 
	\centering
	\begin{tabular}{ |c|c|c|c|}
		\hline \hline 
		    &  D   &    E [Gev]  &   F    \\ \hline
		for $\langle\Theta^{f}\rangle$ &  $0.009$  &    0.040    &    0.024   \\   
		for $\langle\Theta^{g}\rangle$  &  $0.091$  &    0.047    &  -0.731   \\
		\hline \hline 
	\end{tabular}
	\caption{The coefficients D, E and F in Eq. (\ref{tetaf}) .}
	\label{tab:coefficients2}
\end{table}

In the next step, we have to obtain the temperature-dependent continuum threshold $ s_{0}(T) $. For this purpose, we use the following parametrization:
\begin{eqnarray}\label{continuumthreshold}
s_{0}(T)=s_{0} f(T),
\end{eqnarray}
where $ s_0 $ is continuum threshold at zero temperature. It is  not totally arbitrary and depends on the energy of the first excited state lies just above the considered positive and negative parity states with the same quantum numbers. $ s_{0}(T) $  should reduce to $ s_0 $ at $T \rightarrow 0$ limit and $f(T) $ must obey $f(T) \rightarrow 1 $ at this limit. As we mentioned in the previous section, the four unknowns (masses of the positive and negative parity baryons as well as their residues) at each channel are obtained in terms  of Borel parameter, QCD degrees of freedom and  $ s_{0}(T) $.    As it is seen the temperature-dependent continuum threshold, $ s_{0}(T) $ , contains the vacuum continuum threshold and $ f(T) $. The $ s_0 $ together with $ M^2 $ are determined considering the standard prescriptions of the method and will be discussed below.  One of the main tasks is to determine the function $ f(T)  $, which plays an important role in determination of the behavior of the physical quantities with respect to the temperature. To determine $ f(T)  $ we  apply another derivative with respect to    $ \frac{d}{d(-\frac{1}{M^2})} $  to both sides of Eq. (\ref{derivativeresiduesumrule}) and  use the expressions of the masses and residues, obtained in the previous section that also contain $ f(T)  $ though $ s_{0}(T) $, in the resultant equation . By numerical solving of the obtained equation at different temperature (up to thepseudocritical temperature used in the present study), we find the following fit function for $ f(T) $:
\begin{eqnarray}\label{fTcontinuumthreshold}
f(T)=1-0.96\Big(\frac{T}{T_{pc}}\Big)^{9}.
\end{eqnarray}

\begin{figure}[ht]
	\begin{center}
	\includegraphics[width=8cm]{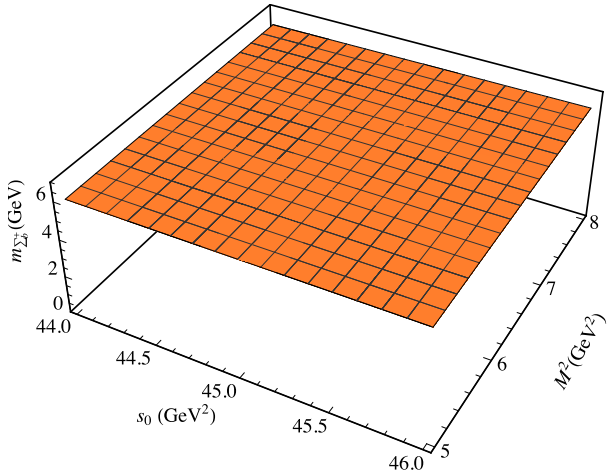}
	\end{center}
	\caption{The mass of the positive parity $\Sigma_{b}$ baryon as  functions of $M^2$ and $ s_{0} $ at $T=0$.} \label{fig1}
\end{figure}

Let us now discuss how we  determine  the working windows  for the  Borel parameter and vacuum continuum threshold $ s_0 $. They are fixed using the standard criteria of the method. Namely, the pole dominance, OPE convergence and weak dependence of the physical quantities on these parameters. All of these requirements lead to the following working intervals for different members of the baryons under study:
\begin{eqnarray}
43.0 ~ GeV^2  \leqslant s^{\Lambda_b}_0 \leqslant 44.0 ~ GeV^2, \nonumber \\
44.0 ~ GeV^2  \leqslant s^{\Sigma_b}_0 \leqslant 46.0 ~ GeV^2,
\nonumber \\
45.0 ~ GeV^2  \leqslant s^{\Xi_b}_0 \leqslant 46.0 ~ GeV^2, \nonumber \\
47.0 ~ GeV^2  \leqslant s^{\Xi_{b}^{'}}_0 \leqslant 48.0 ~ GeV^2,
\nonumber \\
48.5 ~ GeV^2  \leqslant s^{\Omega_{b}}_0 \leqslant 49.5 ~ GeV^2, 
\nonumber \\
M^2\in [5,8]~ GeV^2,~~~~
\end{eqnarray}
for bottom and 
\begin{eqnarray}
8.5 ~ GeV^2  \leqslant s^{\Lambda_c}_0 \leqslant 9.5 ~ GeV^2, \nonumber \\
9.5 ~ GeV^2  \leqslant s^{\Sigma_c}_0 \leqslant 10.5 ~ GeV^2,
\nonumber \\
10.5 ~ GeV^2  \leqslant s^{\Xi_c}_0 \leqslant 11.5 ~ GeV^2, \nonumber \\
11.5 ~ GeV^2  \leqslant s^{\Xi_{c}^{'}}_0 \leqslant 12.5 ~ GeV^2,
\nonumber \\
11.8 ~ GeV^2  \leqslant s^{\Omega_{c}}_0 \leqslant 12.8 ~ GeV^2,
\nonumber \\
M^2\in [3,5]~ GeV^2,~~~~
\end{eqnarray}
for charmed baryons. 
 To check the stability of the results  with respect to the changes in  $ M^{2}$ and $ s_{0} $ in their working intervals, as an example, we plot a 3D graphic of the mass of $\Sigma^0_{b}$ positive parity baryon as functions of these auxiliary parameters at $T=0$ in Figure \ref{fig1}. We see that the dependence of the mass on both $M^2$ and $ s_{0} $ is weak and the changes remain within the acceptable limits of the method.  The parameters of other members show similar behavior.  

Now, we proceed to investigate the temperature dependence of the masses and residues of the $B_{Q}$ baryons. As examples, we only present the results on the temperature-dependent masses and residues  for the positive parity baryons. They reflects behavior of the OPE sides, which are common for both the positive and negative parity baryons, i.e., the masses and residues of both parities are obtained in terms of the functions in OPE sides as presented in the previous section.    To this end, we  plot the ratio of the temperature-dependent  mass (residue) to its  vacuum  value, $ m(T)/m(0) $ ($ \lambda(T)/\lambda(0) $),  for the positive parity baryons as  functions of $M^2$ and ratio $ T/T_{pc}$  in  3D  at average values of $ s_{0}$  for $\Lambda_{b}$, $\Xi_{b}$, $\Sigma_{b}$, $\Xi_{b}^{'}$ and $\Omega_{b}$ baryons in Figs. \ref{fig2} and \ref{fig3}. From these figures, we see that the  spectroscopic parameters of these baryons  remain approximately unchanged with respect to the changes in  $ T/T_{pc}$ up to $T\cong108~ MeV$ for masses  and $T\cong93~ MeV$ for residues.  After these points, they start to decrease rapidly with increasing the temperature and we are witness of melting of these baryons. We realize that similar situation is valid for the charmed  $\Lambda_{c}$, $\Xi_{c}$, $\Sigma_{c}$, $\Xi_{c}^{'}$ and $\Omega_{c}$ baryons.  The amount of negative shifts in masses and residues near to the critical point are shown in Tables \ref{tab:decreaseinmass} and  \ref{tab:decreaseinresidue}. In the case of mass, the order of shifts roughly are comparable between the bottom and charmed baryons of each channel. Among the results,  the order of negative shifts for all channels in  bottom case is roughly the same but shows some differences among the charmed baryons. The sum rules for masses  give reliable predictions up to thepseudocritical point considered in the present study.  As far as the residues are concerned,  the amount of shifts are roughly the same for b- and c-baryons and they are very large. Thus, at   $T\rightarrow T_{pc}$ limit,  the residues approaches to zero and we see the melting of the baryons.

\begin{widetext}

 \begin{table}[ht!] 
	\centering
	\begin{tabular}{ |c|c|c|c|c|c|}
		\hline \hline
		    &$\Lambda_{b(c)}^{+}$ &  $\Xi_{b(c)}^{+}$   & $\Sigma_{b(c)}^{+}$ & $\Xi_{b(c)}^{'+}$ & $\Omega_{b(c)}^{+}$ \\ \hline
		 Negative shift $(\%)$  &  $74(72)$   &    $74(72)$    &    $74(80)$ &   $74(77)$    &   $75(77)$ \\   
\hline \hline
	\end{tabular}
	\caption{At  $T\rightarrow T_{pc}$ limit, the percent  of negative shifts  in masses of the spin-1/2 heavy baryons with the positive parity compared to their vacuum values.}
	\label{tab:decreaseinmass}
\end{table}

\begin{table}[ht!] 
	\centering
	\begin{tabular}{ |c|c|c|c|c|c|}
		\hline \hline
		    &$\Lambda_{b(c)}^{+}$ &  $\Xi_{b(c)}^{+}$   & $\Sigma_{b(c)}^{+}$ & $\Xi_{b(c)}^{'+}$ & $\Omega_{b(c)}^{+}$ \\ \hline
\hline 
Negative shift $(\%)$  &  $92$   &    $94$    &    $96$ &   $97$    &   $96$ \\   
\hline \hline
	\end{tabular}
	\caption{The percent of negative shift  in residues of the spin-1/2 heavy baryons with the positive parity relative to their vacuum values near to thepseudocritical point.}
	\label{tab:decreaseinresidue}
\end{table}
\end{widetext}

\begin{widetext}

	\begin{figure}[ht]
		\begin{center}
		\subfigure[]{\includegraphics[width=8cm]{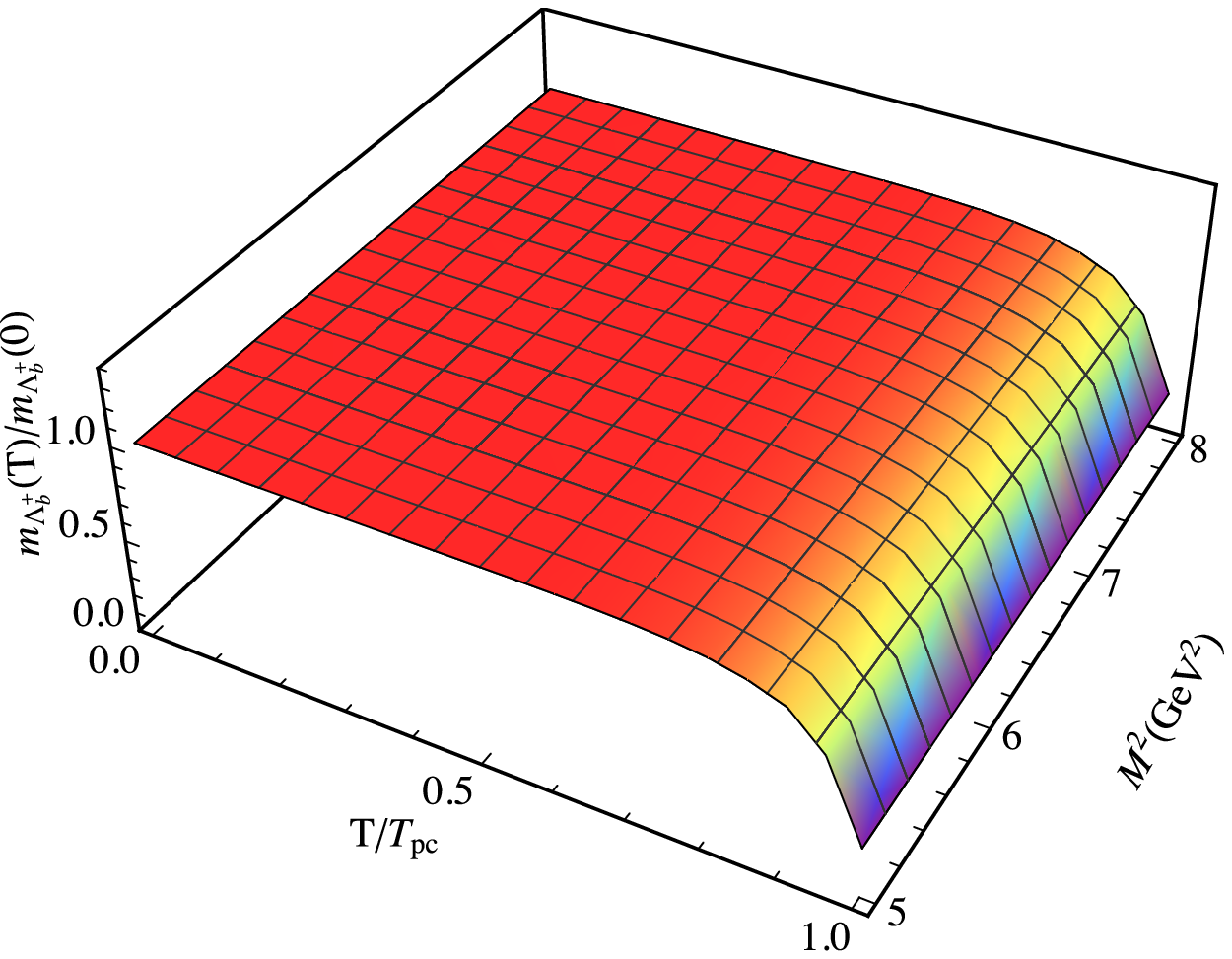}}
		\subfigure[]{\includegraphics[width=8cm]{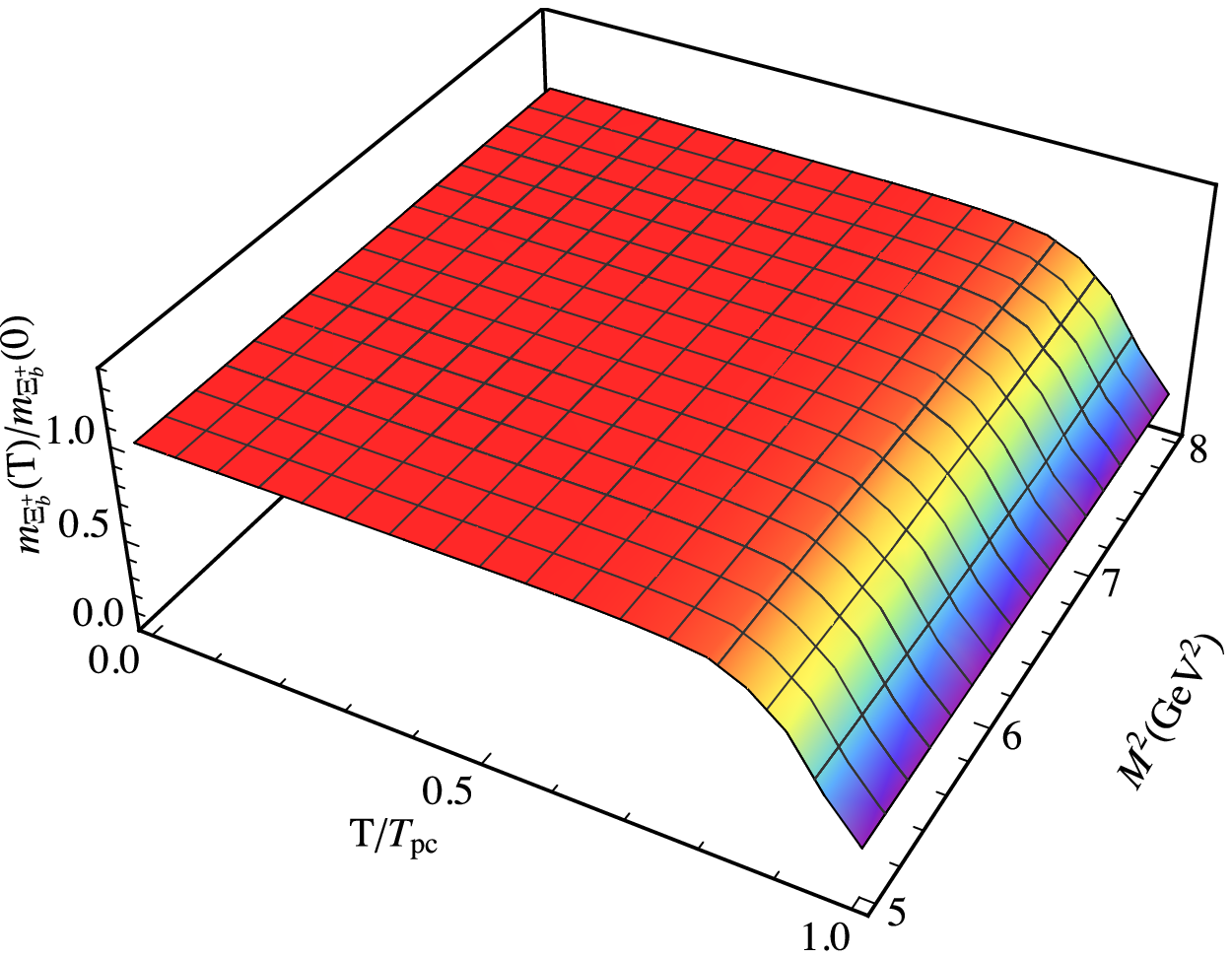}}
			\subfigure[]{\includegraphics[width=8cm]{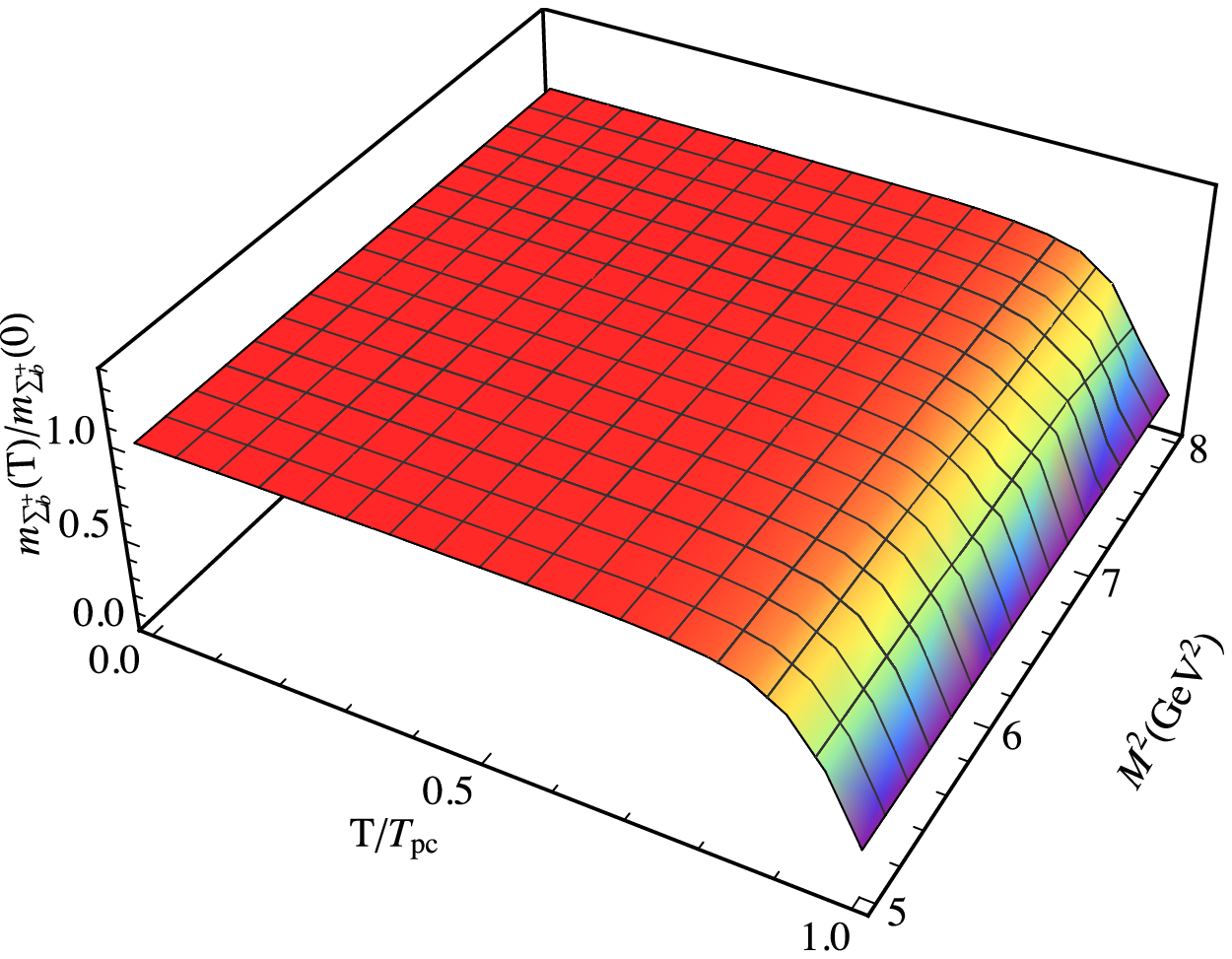}}
		\subfigure[]{\includegraphics[width=8cm]
{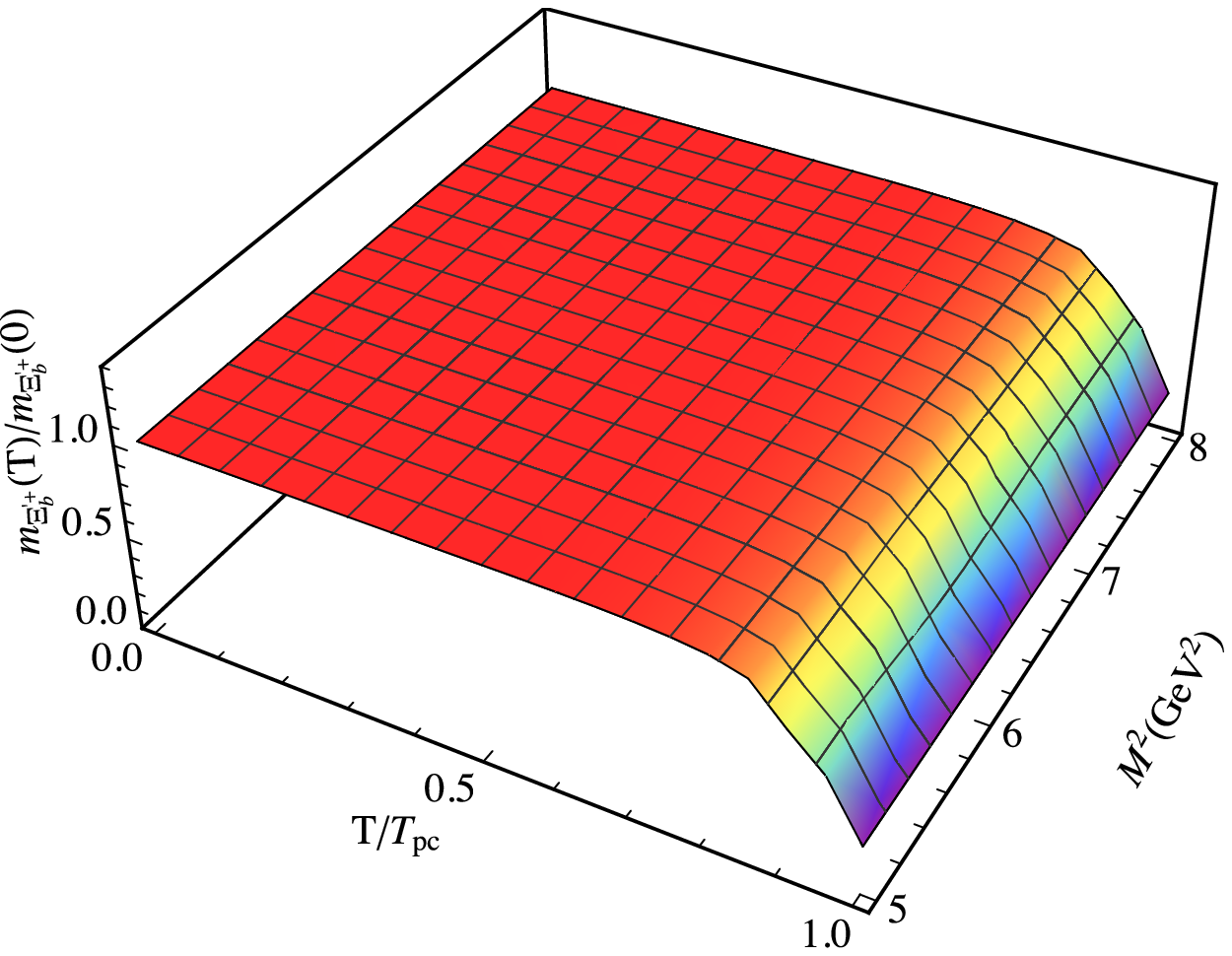}}
		\subfigure[]{\includegraphics[width=8cm]
{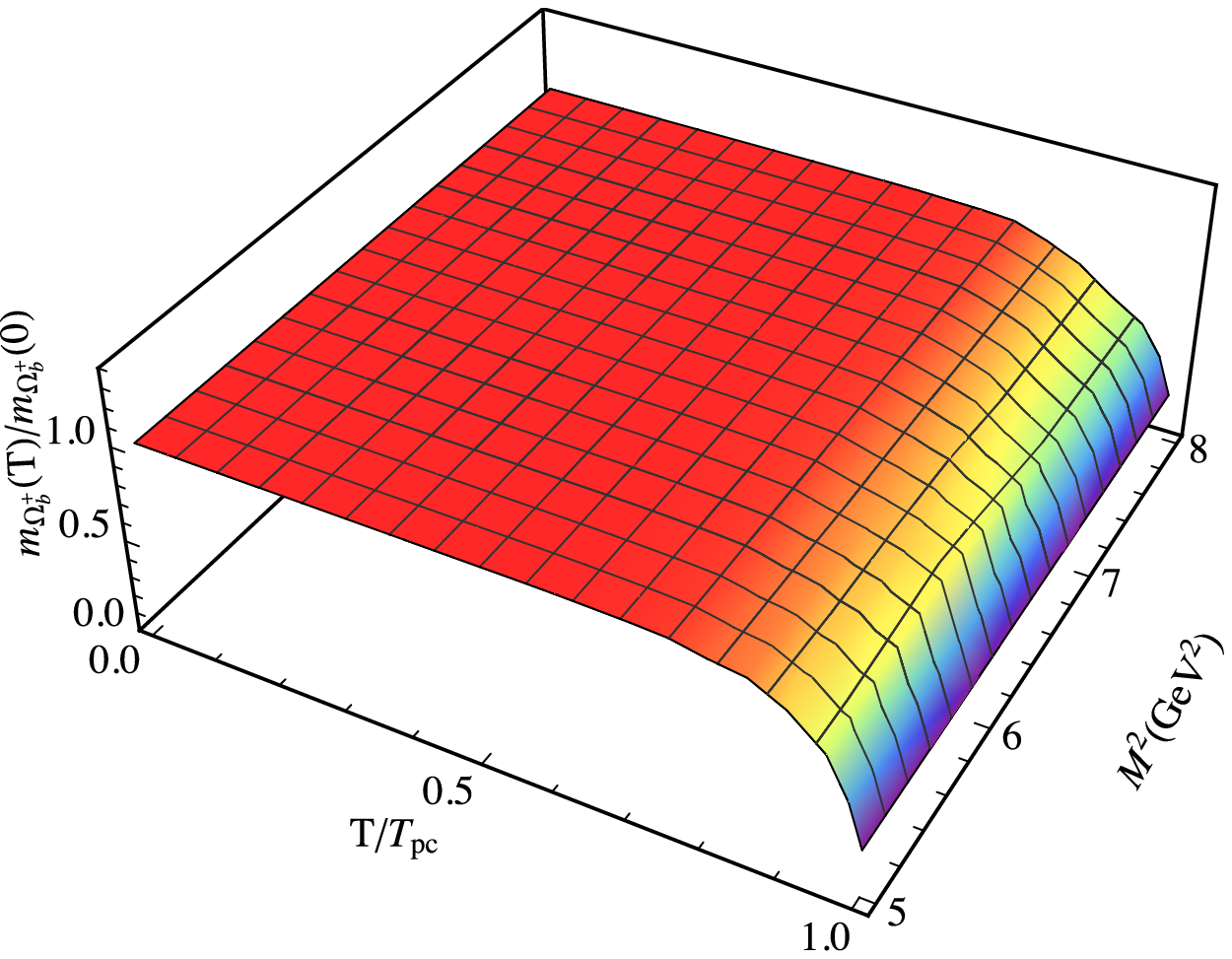}}
		\end{center}
		\caption{The ratio $ m(T)/m(0)$ as  functions of $M^2$ and $T/T_{pc}$ for positive parity $\Lambda_{b}$, $\Xi_{b}$, $\Sigma_{b}$, $\Xi_{b}^{'}$ and $\Omega_{b}$ baryons  at average
value of $s_{0}$.} \label{fig2}
	\end{figure}
	
	\begin{figure}[ht]
		\begin{center}
			\subfigure[]{\includegraphics[width=8cm]{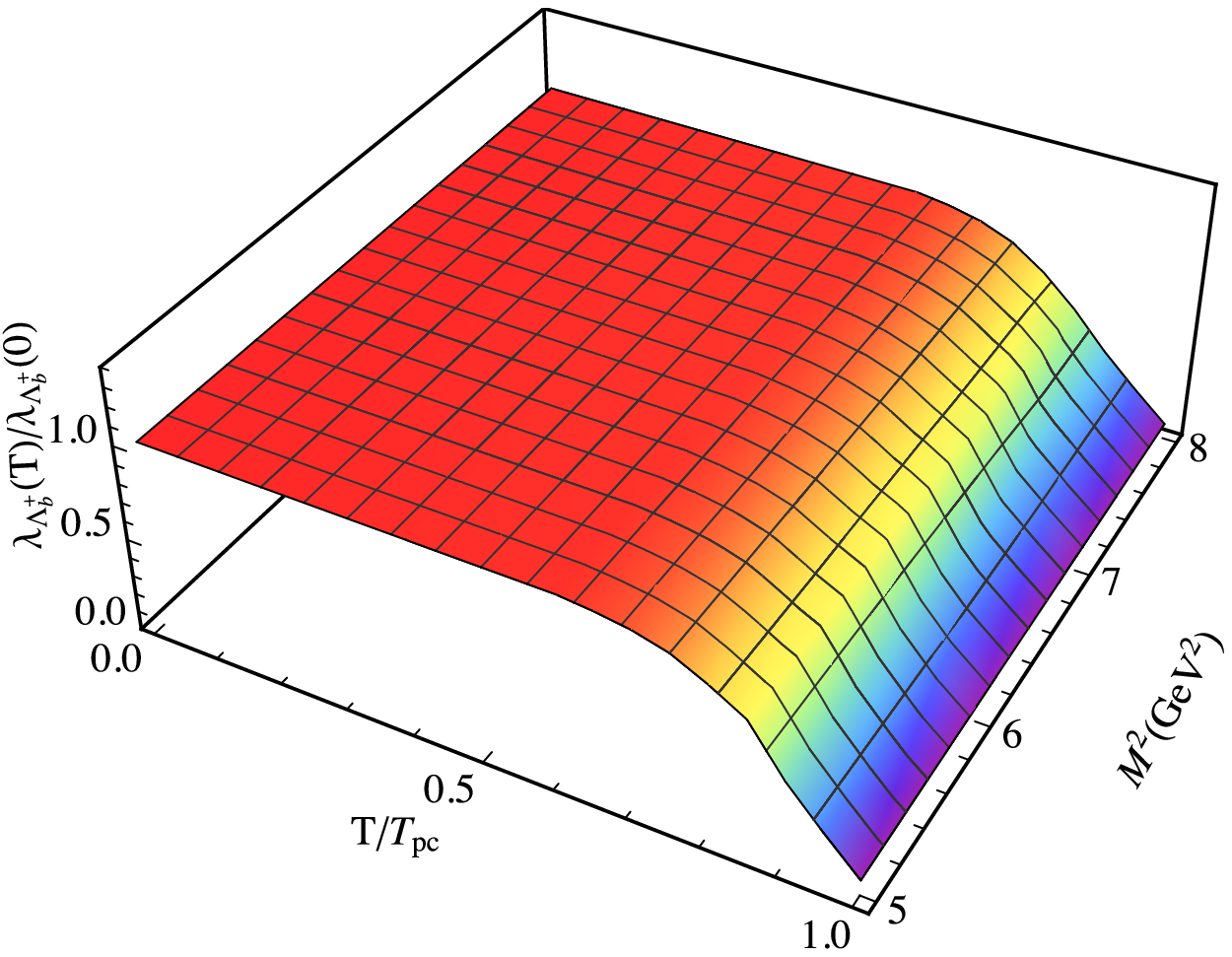}}
			\subfigure[]{\includegraphics[width=8cm]{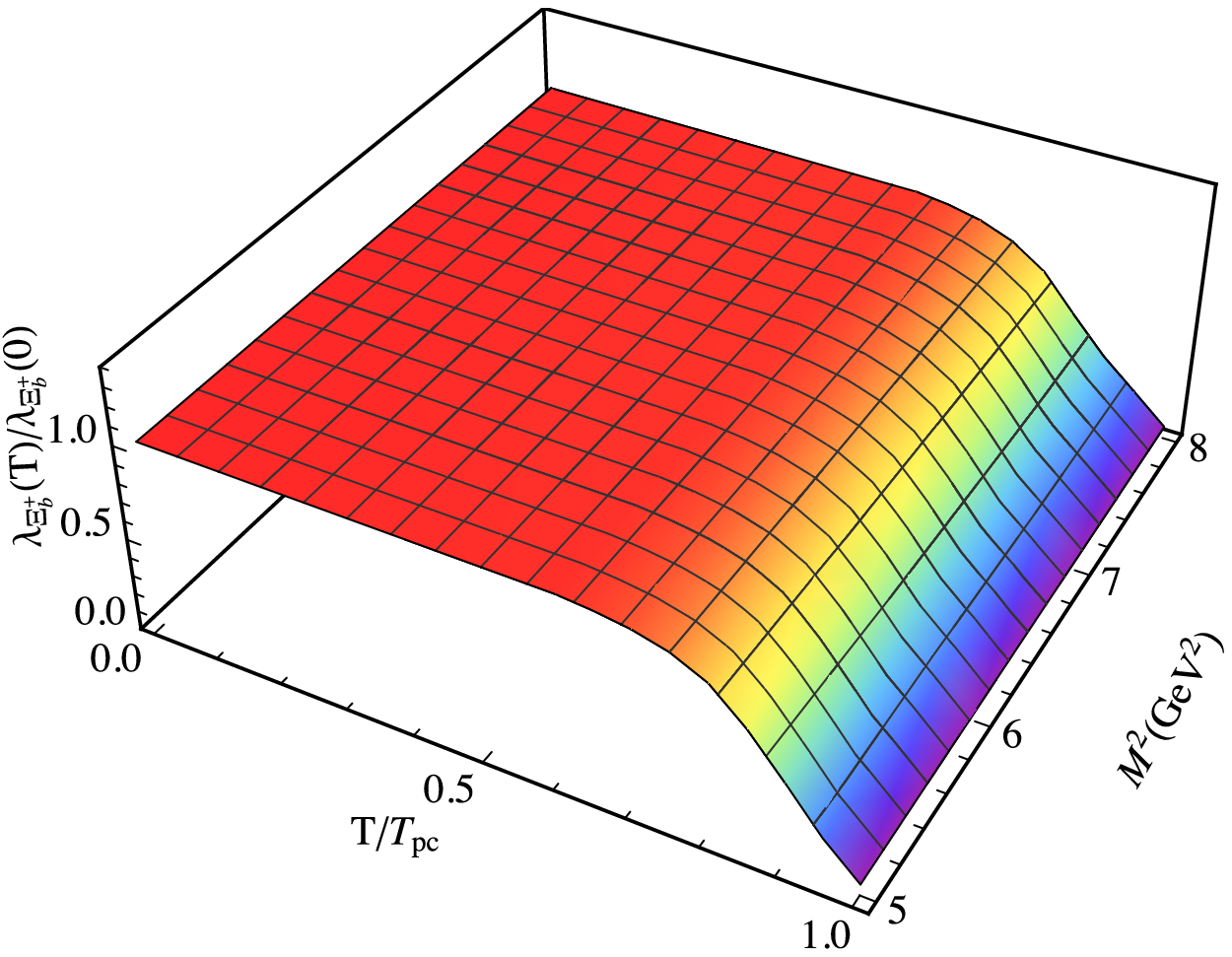}}
		\subfigure[]{\includegraphics[width=8cm]{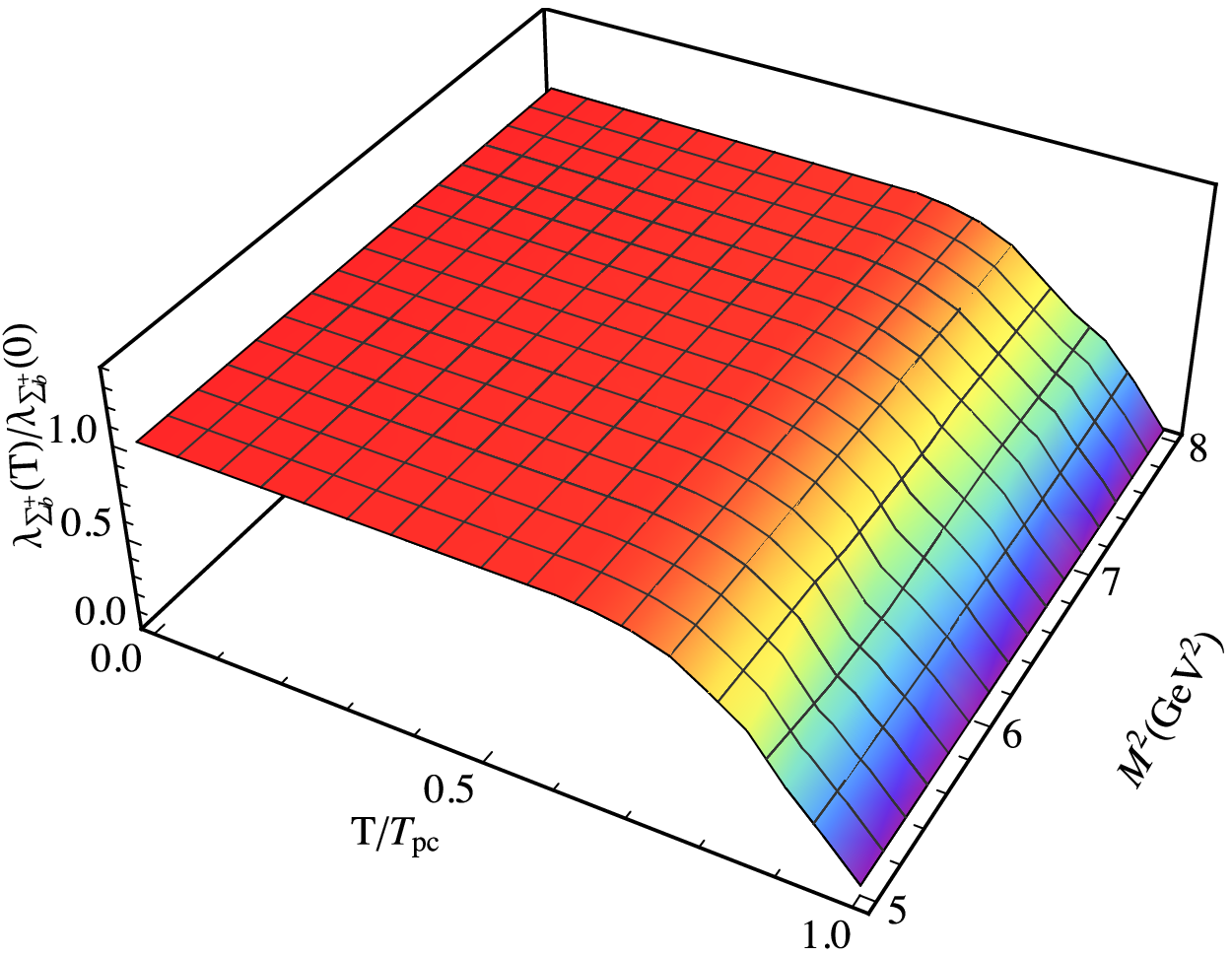}}
		\subfigure[]{\includegraphics[width=8cm]{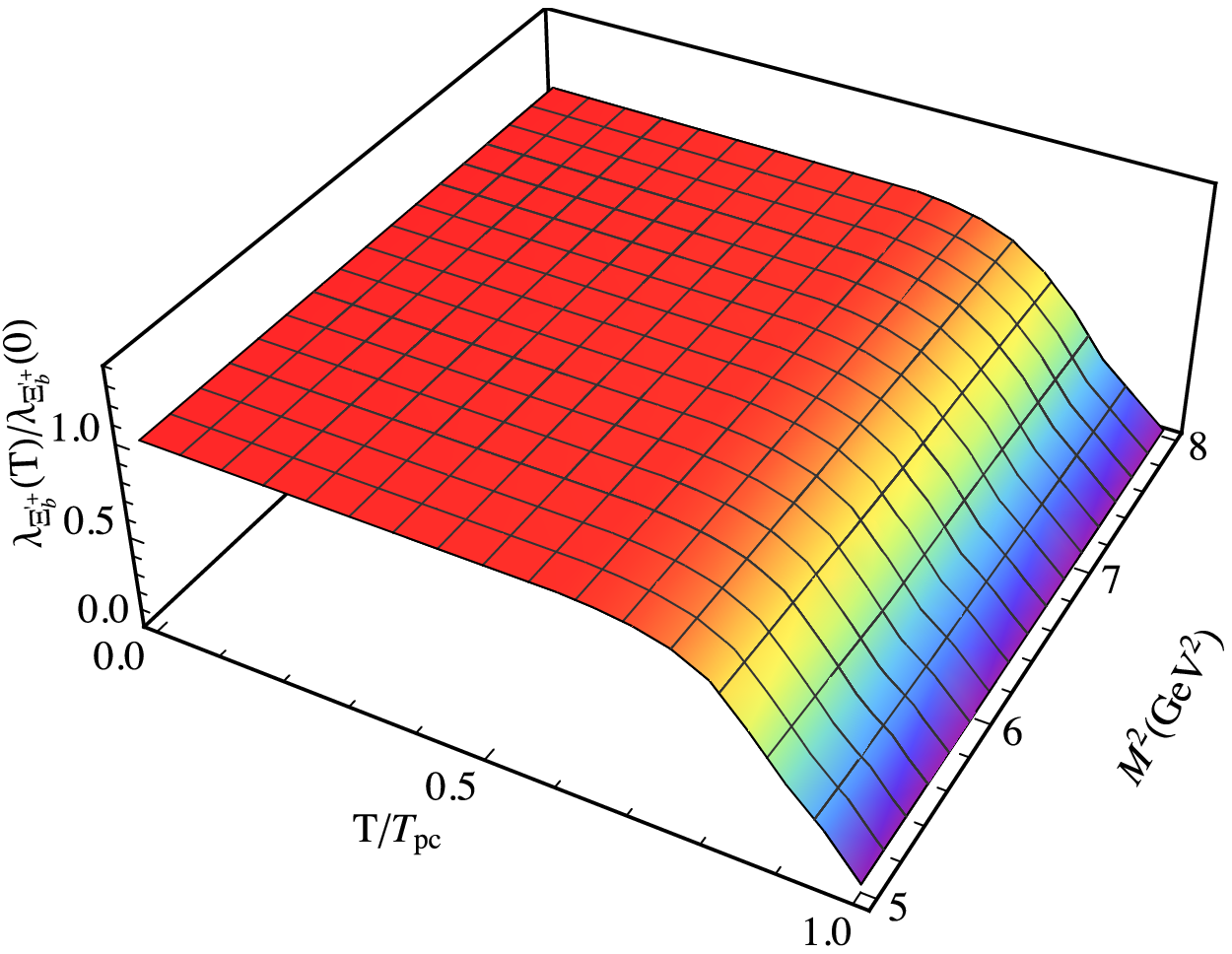}}
		\subfigure[]{\includegraphics[width=8cm]{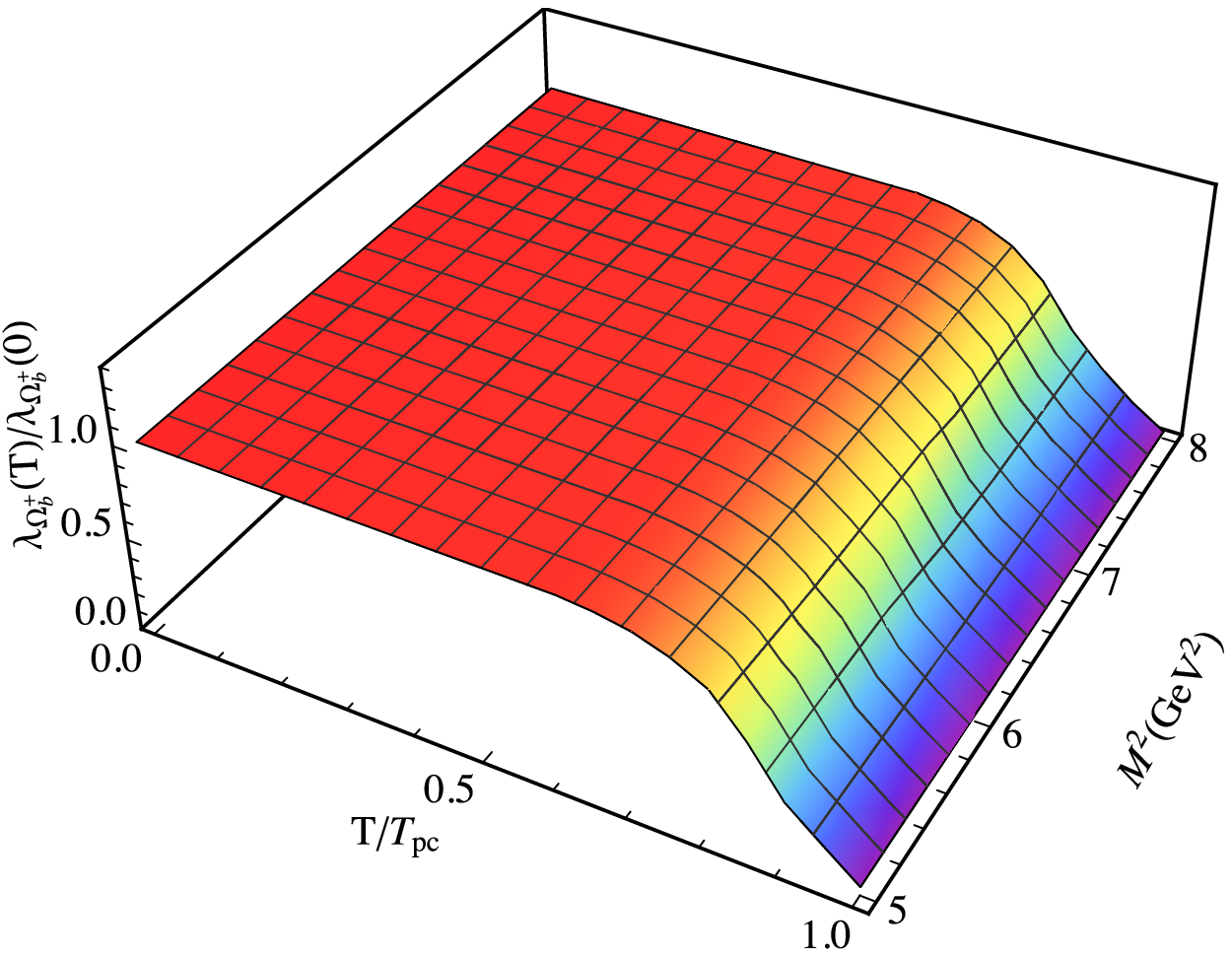}}
		\end{center}
	\caption{The ratio $ \lambda(T)/\lambda(0) $ as  functions of $M^2$ and $T/T_{pc}$ for positive parity  $\Lambda_{b}$, $\Xi_{b}$, $\Sigma_{b}$, $\Xi_{b}^{'}$ and $\Omega_{b}$ baryons at average value of $s_{0}$.} \label{fig3}
	\end{figure}
	\end{widetext}
 
Our final task in this section is to discuss the  results at $ T\rightarrow 0 $ limit. In this limit, the values of masses  for the spin-1/2 heavy  baryons containing a bottom quark with both the positive and negative parities are presented in Tables \ref{tab:1} and \ref{tab:2}, respectively. Similarly, the vacuum masses for the spin-1/2 heavy  baryons containing a charm quark with both the positive and negative parities are presented in Tables \ref{tab:3} and \ref{tab:4}, respectively. From these tables, we see that our results are in good consistency  with existing  experimental data and other  theoretical predictions within the presented uncertainties. These consistency lead us to  hope that, the  obtained results at nonzero temperature in the present study  can shed light on the future heavy ion collision experiments and can be used in analyses of the related data. The uncertainties in our predictions belong to those related to the working intervals of the auxiliary parameters as well as the uncertainties of all the presented input parameters.

\begin{widetext}
	
\begin{table}[h]
\centering
\begin{tabular}{|c||c|c|c|c|c|}\hline
&$m_{\Lambda_{b}^{+}}$ &  $m_{\Xi_{b}^{+}}$   & $m_{\Sigma_{b}^{+}}$ & $m_{\Xi_{b}^{'+}}$ & $m_{\Omega_{b}^{+}}$\\\cline{1-6}
\hline\hline
present work&$5.691^{+0.101}_{-0.109}$ &$5.797^{+0.078}_{-0.077}$&$5.845^{+0.137}_{-0.144}$&$5.957^{+0.147}_{-0.157}$&$6.065^{+0.113}_{-0.112}$\\\cline{1-6}
Exp\cite{PDG}&$ 5.61960\pm0.00017 $&$ 5.7919\pm0.0005 $&$ 5.81056\pm0.00025 $&$ 5.93502\pm0.00002\pm0.00005 $&$ 6.0461\pm0.0017 $\\\cline{1-6}
\cite{Roberts}& 5.612  & 5.844 & 5.833  &$-$&6.081\\\cline{1-6}
\cite{Karliner}&$-$& 5.790-5800 &$-$& $ 5.930\pm0.005 $&$ 6.0521\pm0.0056 $\\\cline{1-6}
\cite{Capstick}&5.585&$-$&5.795&$-$&$-$\\\cline{1-6}
\cite{Dai}&$-$&$-$&$ 5.83\pm0.09 $&$-$&$-$\\\cline{1-6}
\cite{Liu}&$5.637^{+0.068}_{-0.056}$&$5.780^{+0.073}_{-0.068}$&$5.809^{+0.082}_{-0.076}$&$5.903^{+0.081}_{-0.079}$&$6.036\pm0.081 $\\\cline{1-6}
\cite{Korner}&$ 5.641\pm0.05 $& 5.80 & 5.82 & 5.94  & 6.04 \\\cline{1-6}
\cite{Roncaglia}&$ 5.620\pm0.040 $&$ 5.810\pm0.040 $&$ 5.820\pm0.040$&$ 5.950\pm0.040 $&$ 6.060\pm0.050 $\\\cline{1-6}
\cite{Ghalenovi1}&$-$&5.833&5.815&$-$&5.948\\\cline{1-6}
\cite{Ghalenovi2}&5.683  &5.833&5.708&$-$&5.967\\\cline{1-6}
\cite{Brown}&5.626&5.771&5.856&5.933&6.056\\\cline{1-6}
\cite{Mathur}&5.664&5.762&$-$&$-$&6.021\\\cline{1-6}
\cite{Ebert1}&5.622&5.812&5.805&5.937&6.065\\\cline{1-6}
\cite{Ebert2}&5.622&5.812&5.805&$-$&6.065\\\cline{1-6}
\cite{Ebert3}&5.620&5.803&5.808&$-$&6.064\\\cline{1-6}
\cite{Kim}&5.609&5.8036&5.8055&5.9338&6.0571\\\cline{1-6}
\cite{Yin}&5.62 &5.75 &5.75 &5.88 &6.00\\\cline{1-6}
\cite{Azizi}&$ 5.614\pm0.345 $&$-$&$ 5.810\pm0.241 $&$-$&$-$\\\cline{1-6}
\cite{Wang1,Wang2}&$5.618^{+0.078}_{-0.104}$&$-$&$5.96\pm0.10 $&$-$&$-$\\\cline{1-6}
\cite{Wang3}& $5.65\pm0.20 $&$5.73\pm0.18$ &$-$ &$-$&$-$\\\cline{1-6}
\cite{Zhang1,Zhang2}&$ 5.69\pm0.13 $&$ 5.75\pm0.13 $&$ 5.73\pm0.21 $&$ 5.87\pm0.20 $&$ 5.89\pm0.18 $\\\cline{1-6}
\cite{Agaev2}&$-$&$-$&$-$&$-$&6.487$\pm0.187$\\\cline{1-6}
\cite{Agaev4}&$-$&$-$&$-$&$-$&$6.024\pm0.065$\\\cline{1-6}
\end{tabular}
\vspace{0.8cm}
\caption{Vacuum masses of the spin-1/2 positive parity heavy baryons containing a bottom quark  (in $GeV$).
}\label{tab:1}
\end{table}

\begin{table}[h]
\centering
\begin{tabular}{|c||c|c|c|c|c|}\hline
&$m_{\Lambda_{b}^{-}}$ &  $m_{\Xi_{b}^{-}}$   & $m_{\Sigma_{b}^{-}}$ & $m_{\Xi_{b}^{'-}}$ & $m_{\Omega_{b}^{-}}$\\\cline{1-6}
\hline\hline
present work&$5.910^{+0.118}_{-0.132}$ &$6.095^{+0.135}_{-0.150}$&$6.143^{+0.095}_{-0.087}$&$6.255^{+0.113}_{-0.104}$&$6.370^{+0.066}_{-0.061}$\\\cline{1-6}
Exp\cite{PDG}&$ 5.91219\pm0.00017 $&$ - $&$ - $&$ - $&$ - $\\\cline{1-6}
\cite{Roberts}& 5.939  & 6.108 & 6.099  &$-$&6.301\\\cline{1-6}
\cite{Capstick}&5.912&$-$&6.070&$-$&$-$\\\cline{1-6}
\cite{Ebert2}&5.930&6.119&$ 6.108 $&$-$&6.352\\\cline{1-6}
\cite{Ebert3}&$5.930$&$6.120$&$6.101$&$-$&$6.339 $\\\cline{1-6}
\cite{Garcilazo}&$ 5.890 $& 6.076 & 6.039 &$ -$  & 6.278 \\
\cline{1-6}
\cite{Yin}&$ 6.03 $&$ 6.15 $&$6.32$&$6.40$&$6.49$\\\cline{1-6}
\cite{Wang3}&$ 5.85\pm0.18 $&$ 6.01\pm0.16 $&$-$&$-$&$-$\\\cline{1-6}
\cite{Zhang1,Zhang2}&$ 5.85\pm0.15 $&$ 5.95\pm0.16 $&$ - $&$ - $&$ - $\\\cline{1-6}
\cite{Agaev2}&$ - $&$ - $&$ -$&$ - $&$ 6.336\pm0.183 $\\\cline{1-6}
\end{tabular}
\vspace{0.8cm}
\caption{Vacuum masses of the spin-1/2 negative parity heavy baryons containing a bottom quark (in $GeV$).
}\label{tab:2}
\end{table}

\begin{table}[h]
\centering
\begin{tabular}{|c||c|c|c|c|c|}\hline
 & $m_{\Lambda_{c}^{+}}$ &$m_{\Xi_{c}^{+}}$  & $m_{\Sigma_{c}^{+}}$& $m_{\Xi_{c}^{'+}}$& $m_{\Omega_{c}^{+}}$\\\cline{1-6}
\hline\hline
present work&$2.283^{+0.087}_{-0.095}$ &$2.460^{+0.025}_{-0.094}$&$2.488^{+0.105}_{-0.113}$&$2.576^{+0.095}_{-0.101}$&$2.689^{+0.100}_{-0.093}$\\\cline{1-6}
Exp\cite{PDG}&$ 2.28646\pm0.00014 $&$ 2.46771\pm0.00023$&$ 2.4529\pm0.0004 $&$ 2.5787\pm0.0005 $&$ 2.6952\pm0.0017 $\\\cline{1-6}
\cite{Roberts} & 2.268 & 2.492  & 2.455 &$-$&2.718\\\cline{1-6}
\cite{Capstick}&2.265&$-$&2.440&$-$&$-$\\\cline{1-6}
\cite{Dai}&$-$&$-$&$ 2.52\pm 0.08$&$ 2.5808 \pm 0.0021$&$-$\\\cline{1-6}
\cite{Liu}&$2.271^{+0.067}_{-0.049}$&$2.432^{+0.079}_{-0.068}$&$2.411^{+0.093}_{-0.081}$&$2.508^{+0.097}_{-0.091}$&$2.657^{+0.102}_{-0.099}$\\\cline{1-6}
\cite{Korner}&$ 2.285\pm0.0006 $ &$ 2.4728\pm0.0017 $&$ 2.4525\pm0.0009 $ & 2.57 &$ 2.719\pm0.007\pm0.0025 $\\\cline{1-6}
\cite{Roncaglia}&$ 2.285\pm0.001 $&$ 2.468\pm0.003 $&$ 2.453\pm0.003 $&$ 2.580\pm0.020 $&$ 2.710\pm0.030 $\\\cline{1-6}
\cite{Ghalenovi1}&$-$&2.473&2.455&$-$&2.588\\\cline{1-6}
\cite{Ghalenovi2}&2.303 &2.453&2.328&$-$&2.587\\\cline{1-6}
\cite{Patel1}&$-$&2.653&2.586&$-$&2.720\\\cline{1-6}
\cite{Patel2}&$-$&2.648&2.575&$-$&2.723\\\cline{1-6}
\cite{Brown}&2.254&2.433&2.474&2.574&2.679\\\cline{1-6}
\cite{Mathur}&$-$&2.440&2.407&$-$&2.652\\\cline{1-6}
\cite{Lewis}&2.295&2.462&2.490&2.594&2.699\\\cline{1-6}
\cite{Bahtiyar}&2.343(23)& 2.474(11) &$ 2.459(45) $&$2.593(22)$&2.711(16)\\\cline{1-6}
\cite{Ebert1,Ebert2}&2.297&2.481&2.439&2.578&2.698\\\cline{1-6}
\cite{Ebert3}&2.286&2.476&2.443&$-$&2.698\\\cline{1-6}
\cite{Kim}&2.2807&2.4752&2.4485&2.5768&2.7001\\\cline{1-6}
\cite{Yin}&2.40&2.55&2.45&2.59&2.73\\\cline{1-6}
\cite{Azizi}&$ 2.295\pm0.251 $&$-$&$ 2.451\pm0.208 $&$-$&$-$\\\cline{1-6}
\cite{Wang1,Wang2}&$2.284^{+0.049}_{-0.078}$&$-$&$2.54\pm0.15 $&$-$&$-$\\\cline{1-6}
\cite{Wang3}&$ 2.26\pm0.27 $&$2.44\pm0.23$ &$ - $&$-$&$-$\\\cline{1-6}
\cite{Zhang1,Zhang2}&$ 2.31\pm0.19 $&$ 2.48\pm0.21 $&$ 2.40\pm0.31 $&$ 2.50\pm0.29 $&$ 2.62\pm0.29 $\\\cline{1-6}
\cite{Agaev1}&$ - $&$-$ &$ - $&$2.925\pm0.115$ &$-$\\\cline{1-6}
\cite{Agaev3,Agaev4}&$-$&$-$ & $-$ &$-$&$2.685\pm0.123$ \\\cline{1-6}
\end{tabular}
\vspace{0.8cm}
\caption{Vacuum masses of the spin-1/2 positive parity heavy baryons containing a charm quark (in $GeV$).   
}\label{tab:3}
\end{table}

\begin{table}[h]
\centering
\begin{tabular}{|c||c|c|c|c|c|}\hline
 & $m_{\Lambda_{c}^{-}}$ &$m_{\Xi_{c}^{-}}$  & $m_{\Sigma_{c}^{-}}$& $m_{\Xi_{c}^{'-}}$& $m_{\Omega_{c}^{-}}$\\\cline{1-6}
\hline\hline
present work&$2.548^{+0.090}_{-0.086}$ &$2.763^{+0.103}_{-0.102}$&$2.868^{+0.051}_{-0.037}$&$2.901^{+0.082}_{-0.079}$&$3.099^{+0.046}_{-0.024}$\\\cline{1-6}
Exp\cite{PDG}&$ 2.59225\pm0.00028 $&$ 2.7919\pm0.0005$&$ - $&$ - $&$ - $\\\cline{1-6}
\cite{Roberts}& 2.625  & 2.763 & 2.748  &$-$&2.977\\\cline{1-6}
\cite{Capstick}&2.630&$-$&2.795&$-$&$-$\\\cline{1-6}
\cite{Bahtiyar}&2.668(16)& 2.770(67) &$ 2.814(20) $&$2.933(16)$&3.044(15)\\\cline{1-6}
\cite{Ebert2}&2.598& 2.801 &$ 2.795 $&$-$&3.020\\\cline{1-6}
\cite{Ebert3}&$2.598 $&$2.792 $&$ 2.799 $&$-$&$3.055 $\\\cline{1-6}
\cite{Migura}&$ 2.594 $& 2.769 & 2.769 & $-$  &$ -$ \\
\cline{1-6}
\cite{Gerasyuta1,Gerasyuta2,Gerasyuta3}&$ 2.400 $& $-$ & 2.700 & $-$  & $-$ \\
\cline{1-6}
\cite{Garcilazo}&$ 2.559 $& 2.749 & 2.706 & $-$  & 2.959 \\
\cline{1-6}
\cite{Yin}&$ 2.67 $& 2.79 & 2.84 & $2.94$  & 3.03 \\
\cline{1-6}
\cite{Wang3}&$ 2.61\pm0.21 $&$ 2.76\pm0.18 $&$-$&$-$&$-$\\\cline{1-6}
\cite{Zhang1,Zhang2}&$ 2.53\pm0.22 $&$ 2.65\pm0.27 $&$ - $&$ - $&$ - $\\\cline{1-6}
\cite{Agaev1}&$ - $&$ - $&$ -$&$ 2.925\pm0.115   $&$ - $\\\cline{1-6}
\cite{Agaev3,Agaev4}&$ - $&$ - $&$ -$&$ - $&$ 2.990\pm0.129 $\\\cline{1-6}
\end{tabular}
\vspace{0.8cm}
\caption{Vacuum masses of the spin-1/2 negative parity heavy baryons containing a charm quark  (in $GeV$).   
}\label{tab:4}
\end{table}

\end{widetext}

\label{sec:results}
\section{Summary and Conclusions}

We investigated  the mass and residue of the  spin-1/2 single heavy $\Lambda_{Q}$, $\Xi_{Q}$, $\Sigma_{Q}$, $\Xi_{Q}^{'}$ and $\Omega_{Q}$ baryons  as functions of  temperature in the framework of  thermal QCD sum rule. In our calculations, we took into account the non-perturbative operators up to mass dimension 8 including those arising   from the Wilson expansion at finite temperature  due to breaking  the Lorentz invariance.  The obtained results indicate that the mass of these baryons in both the bottom and charm channels remain stable up to roughly  $T=108$ MeV while their residue are unchanged up to $T=93$ MeV. After these points, the masses and residues start to diminish by increasing in the temperature. The shifts in the  mass and residue for both the bottom and charm channels are considerably large and we observe the melting of these baryons near to thepseudocritical temperature determined by recent lattice QCD calculations. The amount of negative shifts near to thepseudocritical point have been  shown in Tables \ref{tab:decreaseinmass} and  \ref{tab:decreaseinresidue}. The order of shifts in masses are  roughly the same between the bottom and charmed baryons of each channel. Among the results,  the order of shifts for all channels for bottom baryons is roughly the same but shows some differences among the charmed members. The sum rules for masses  give reliable predictions up to thepseudocritical point considered in the present study (roughly $ 155 $ MeV).  As far as the residues are concerned,  the amounts of shifts in bottom and charmed cases of each channel are the same. The  negative shifts near to the end point are very large for all baryons and the residues approach to zero atpseudocritical point.

We presented our results for the mass of the single heavy  baryons with  both the positive and negative parities at  $ T\rightarrow 0 $ limit in Tables \ref{tab:1}, \ref{tab:2}, \ref{tab:3} and \ref{tab:4}. We observed that the obtained results for the single heavy bottom and charmed baryons of spin-1/2  with both the positive and negative parities in the present study are in good consistencies with the experimental data presented in  Ref. \cite{PDG}  as well as  with other theoretical predictions made via different phenomenological approaches. Our results on the behavior of the physical quantities considered in the present study with respect to temperature and the amount of shifts in these quantities near to the pseudocritical point may be checked via other phenomenological approaches. The obtained results in the present study may  shed light on analyses of the data provided by the future heavy ion collision experiments.


  \section{Appendix}

In this appendix, we present the explicit forms of the spectral density  $\rho^{ QCD}_{1}(s,T)$ (for perturbative and some non-perturbative parts) and the function $\hat{B}\Gamma_{1}^{QCD}(T)$  defining other non-perturbative contributions for  $ \Sigma^0_{b}$ baryon as examples.  They are obtained as
\begin{widetext}
\begin{eqnarray}\label{RhoPert}
\rho_{1}^{\mathrm{pert.}}(s,T)&=& -\frac{1}{64 \pi^4 }\int_{0}^{1}\frac{dz}{(z-1)} \Bigg\{6\beta \Big[m_{b}^2 m_{u}m_{d}-\beta s m_{u}m_{d} \Big]-3z^{2} m_{b}^4 - 8\beta z^{2} s m_{b}^{2}
-5\beta^{2} z^{2}s^{2}\Bigg\}\times \Theta[L(s,z)],
\end{eqnarray}
\begin{eqnarray}\label{Rhoqbarq}
\rho_{1}^{\langle q\bar{q}\rangle}(s,T)&=& -\frac{1}{8 \pi^2}\int_{0}^{1} dz \beta \Bigg\{m_{u}\Big[\langle
\bar{d}d\rangle- 3z \langle
\bar{u}u\rangle \Big]
+ m_{d}\Big[\langle
\bar{u}u\rangle - 3z \langle
\bar{d}d\rangle \Big]\Bigg\}\times \Theta[L(s,z)],
\end{eqnarray}
\begin{eqnarray}\label{RhoDim4axial}
&&\rho_{1}^{\langle G^2 \rangle+\langle\Theta^{f,g}\rangle}(s,T)=\frac{1}{\pi^2}\int_{0}^{1}dz\Bigg\{z^{2}\Bigg[\frac{g_{s}^{2}}{96\pi^2}\langle
u\Theta^{g}u\rangle-\frac{3}{32}\Big\langle\frac{ \alpha_s G^2}{\pi}\Big\rangle\Bigg]
-\frac{z\beta}{3}\langle
u\Theta^{f}u\rangle\Bigg\}\times \Theta[L(s,z)],
\end{eqnarray}
Here, $z$ is Feynman parameter, $\Theta$ indicates the unit-step function, $L(s,z)=s~z(1-z)-z~m_{b}^2$ and $ \beta=z-1$.
The explicit form of the function $\hat{B}\Gamma_{1}^{QCD}(T)$ for the  $ \Sigma^0_{b}$  baryon is obtained as
\begin{eqnarray}
&&\hat{B}\Gamma_{1}^{QCD}(T)=\frac{1}{1728 \pi^2 M^{6}} \int_{0}^1 dz \frac{e^{\frac{m_{b}^2}{M^{2} \beta}}}{ \beta^{6}} \Bigg\lbrace
\langle \frac{\alpha_s G^2}{\pi} \rangle \Bigg(-9 M^{4} m_{u} m_{d} m_{b}^2 z \beta^{4}-3\beta \pi^2  m_{u} m_{d} m_{b} \langle
\bar{d}d\rangle \Big[m_{b}^2 \beta^{3}\nonumber \\
&+&3 M^{2}\beta^{4}
+M^{2}\beta^{2}(1-4z+3z^{2})\Big]+m_{u}\langle
\bar{u}u\rangle \pi^2 \Big[-63 M^{2}\beta^{2}m_{b}^2(1-z^{3})+180 M^{4}\beta^{6}+9m_{b}^{4}\beta^{3}z\nonumber \\
&-&27 M^{2}m_{b}^{2}\beta(1+3z^{3}-3z^{2})\Big]+m_{d}\langle
\bar{d}d\rangle \pi^2 \Big[M^{2}\beta^{2}m_{b}^2(-36++153z-198z^{2}+81z^{3})+108M^{4}\beta^{4}(1+z^{2})\nonumber \\
&+&8505M^{2}\beta m_{b}^{2}z^{2}+72 M^{4}\beta^{3}(z^{3}-1-6z\beta)+15 m_{b}^{4}z\beta^{3}+45 M^{2}\beta^{4} m_{b}^{2}z\Big]-12 \pi^2 M^{2}\beta^{4}m_{b}^{2}\Big(m_{u}\langle
\bar{d}d\rangle +m_{d}\langle
\bar{u}u\rangle \Big)\nonumber \\
&+&\Big(12 \pi^{2}M^{2}\beta^{3}z m_{b}\langle
\bar{d}d\rangle -6\pi^{2}M^{2}\beta^{3}z m_{b}\langle
\bar{u}u\rangle \Big) \Big(\beta m_{b}^{2}+5M^{2}\beta +2M^{2}\Big)\Bigg)-\Big(12M^{2}\beta^{5}g_{s}^{2} m_{b}^{2} \langle
u\Theta^{g}u\rangle +36 M^{4} \beta^{6}g_{s}^{2} \langle
u\Theta^{g}u\rangle \nonumber \\
&+&48 M^{2}\beta^{6} q_{0}^{2} g_{s}^{2} \langle
u\Theta^{g}u\rangle \Big) \Big(m_{u}\langle
\bar{u}u\rangle +m_{d}\langle
\bar{d}d\rangle \Big)
+\langle
u\Theta^{f}u\rangle \Bigg[\langle \frac{\alpha_s G^2}{\pi} \rangle \Bigg(M^{2}\pi^{2}\beta^{2}m_{b}^{2}(144-176z-80z^{3}+208z^{2})\nonumber \\
&+&96M^{2}\pi^{2}\beta \Big( m_{b}^{2}(1+3z^{3}-3z^{2}-z^{4})-M^{2}\beta^{5}\Big) 
+32m_{b}^{4}z(\pi^{2}+3z-z^{3}+3z^{2})-128\pi^{2}\beta^{4}m_{b}^{2}q_{0}^{2}z \nonumber \\
&+&144M^{4}\pi^{2}\beta^{3}z^{2}(2-\beta)+48M^{4}\pi^{2}\beta^{3} (2-3\beta -6z)\Bigg)
+M^{2}\beta^{3} g_{s}^{2}\langle u\Theta^{g}u\rangle \Big(80 M^{2}\beta^{3} -416M^{2}\beta z+16\beta^{2} m_{b}^{2}+256 \beta^{3} q_{0}^{2}\nonumber \\
&-&96\beta m_{b}^{2}-1536\beta^{2} q_{0}^{2}-96 m_{b}^{2}-1536 \beta q_{0}^{2} \Big)\Bigg]\Bigg\rbrace \Theta[L(s_{0},z)]
+\frac{e^{-\frac{m_{b}^2}{M^{2}}}}{864\pi ^2 M^{8} }\Bigg\lbrace 27 M^{8}m_{0}^{2} \Big(m_{u}\langle
\bar{d}d\rangle
+m_{d}\langle
\bar{u}u\rangle \Big)\nonumber \\
&+&144 \pi^{2}M^{8}\langle
\bar{u}u\rangle \langle
\bar{d}d\rangle 
+72\pi^{2}M^{4} \Big[m_{d}\langle
\bar{d}d\rangle \Big(m_{u}\langle
\bar{u}u\rangle(m_{b}^{2}+M^{2})+M^{2}\langle
u\Theta^{f}u\rangle \Big)
- m_{b}^{2}m_{u}\langle
\bar{u}u\rangle \langle
u\Theta^{f}u\rangle - M^{2} m_{0}^{2}\langle
\bar{u}u\rangle \langle
\bar{d}d\rangle \Big] \nonumber \\
&+&3\pi^{2}M^{4} \langle \frac{\alpha_s G^2}{\pi} \rangle \langle
\bar{u}u\rangle \Big(m_{b}^{2}(m_{u}-m_{d})
-M^{2} (m_{u}+m_{d})\Big)
+48 \pi^{2}M^{6} m_{b} \langle
u\Theta^{f}u\rangle \Big(\langle
\bar{u}u\rangle -\langle
\bar{d}d\rangle \Big)\nonumber \\
&+&9 \pi^{2}M^{4} m_{u}\langle
\bar{u}u\rangle \langle u\Theta^{f}u\rangle \Big(24 M^{2}+64 q_{0}^{2}\Big)-\pi^{2}M^{4} m_{d}\langle
\bar{d}d\rangle \langle u\Theta^{f}u\rangle \Big(120m_{b}^{2}-384q_{0}^{2}\Big) 
-8\pi^{2}m_{0}^{2}m_{b}^{2}\langle
\bar{u}u\rangle \langle
\bar{d}d\rangle \Big(m_{b}^{2}m_{u} m_{d}+9M^{4}\Big)\nonumber \\
&+&128\pi^{2}M^{4}\langle
u\Theta^{f}u\rangle^{2} \Big(m_{b}^{2}-3M^{2}-8q_{0}^{2}\Big)\Bigg\rbrace,
\end{eqnarray}
where the dimensions of some operators included in the formulas are given in Table \ref{tab:dimensions}.
\begin{table}[ht!] 
	\centering
	\begin{tabular}{ |c|c|}
		\hline \hline 
		 Dimension   &  Operator   \\ \hline
		1 &  $I$    \\   
		3 &  $\langle\bar{q}_{1(2)}q_{1(2)}\rangle$     \\
		4 &  $ \langle\Theta^{f(g)}\rangle$     \\
		4 &  $ \langle G^2\rangle $     \\
        5 &  $m_{0}^{2}\langle\bar{q}_{1(2)}q_{1(2)}\rangle$     \\	
	6 &  $\langle\bar{q}_{1}q_{1}\rangle \langle\bar{q}_{2}q_{2}\rangle$     \\	
	6 &  $\langle\bar{q}_{1(2)}q_{1(2)}\rangle^{2}$     \\
	7 &  $\langle\bar{q}_{1(2)}q_{1(2)}\rangle \langle G^2\rangle $     \\
	7 &  $\langle\bar{q}_{1(2)}q_{1(2)}\rangle \langle\Theta^{f(g)}\rangle $     \\
	8 &  $\langle \Theta^{f(g)}\rangle^2 $     \\
	8 &  $ \langle G^2\rangle^2 $     \\
	8 &  $ \langle \Theta^{f(g)}\rangle \langle G^2\rangle  $     \\
		\hline \hline 
	\end{tabular}
	\caption{List of some operators and their mass dimensions entering our calculations.}
	\label{tab:dimensions}
\end{table}
 \end{widetext}



%
%

\end{document}